\documentclass{aa}      
           
\usepackage{graphicx}       
\usepackage{txfonts}        
\usepackage{lscape}
        
\begin{document}        
        
\def\ang{{\rm\AA}}        
        
\def\n{\noindent}        
        
\def\ph{\phantom}         
       
\newcommand{\hei}{He~{\sc i} $\lambda$4471}       
\newcommand{\heii}{He~{\sc ii}  $\lambda$4542}       
\newcommand{\heib}{He~{\sc i} $\lambda$4388}       
\newcommand{\heic}{He~{\sc i} $\lambda$4713}       
\newcommand{\heiib}{He~{\sc ii}  $\lambda$4200}       
\newcommand{\heiic}{He~{\sc ii}  $\lambda$4686}       
\newcommand{\ciii}{C~{\sc iii} $\lambda 1176$}       
\newcommand{\ciiifar}{C~{\sc iii} $\lambda 977$}       
\newcommand{\civ}{C~{\sc iv} $\lambda\lambda$1548,1551}        
\newcommand{\nv}{N~{\sc v}  $\lambda$1239,1243}        
\newcommand{\niv}{N~{\sc iv}  $\lambda$1718}        
\newcommand{\oiv}{O~{\sc iv} $\lambda\lambda$1339,1343}        
\newcommand{\ov}{O~{\sc v}  $\lambda$1371}        
\newcommand{\ovi}{O~{\sc vi} $\lambda\lambda$1032,1038}      
\newcommand{\pv}{P~{\sc v} $\lambda\lambda$1118,1128}        
\newcommand{\siiv}{Si~{\sc iv} $\lambda\lambda$1394,1403}

\title{Analysis of Galactic late-type O dwarfs:\\   
more constraints on the weak wind problem\thanks{Based on observations made with the NASA-CNES-CSA {\sl Far Ultraviolet  
           Spectroscopic Explorer\/} and by the NASA-ESA-SERC {\sl International  
           Ultraviolet Explorer\/}, and retrieved from the Multimission Archive at  
           the Space Telescope Science Institute (MAST). Based on observations collected with the   
           ELODIE spectrograph on the 1.93-m telescope (Observatoire de Haute-Provence, France).  
           Based on observations collected with the FEROS instrument on the  
           ESO 2.2 m telescope, program 074.D-0300 and
           075.D-0061. }$^{,}$\thanks{The Appendix is only available in
           electronic format.} }  
       
\author{W. L. F. Marcolino\inst{1}, J.-C. Bouret\inst{1}, F. Martins\inst{2},   
  D. J. Hillier\inst{3}, T. Lanz\inst{4}, C. Escolano\inst{1}}         
        
\offprints{wagner.marcolino@oamp.fr}        
        
\institute{LAM-UMR 6110, CNRS \& Univ. de Provence, 38 rue Fr\'ederic  Joliot-Curie, F-13388 Marseille, France    
          \and    
           GRAAL-UMR 5024, CNRS \& Univ. de Montpellier II, Place Bataillon, F-34095 Montpellier, France   
           \and   
           Department of Physics and Astronomy, University of Pittsburgh, Pittsburgh, PA 15260, USA   
           \and   
           Department of Astronomy, University of Maryland, College Park, MD 20742, USA   
         }        
        
\date{Received ; Accepted }        
   
\abstract{}         
{To investigate the stellar and wind properties of a sample of late-type O dwarfs          
in order to address the {\it weak wind problem}.}       
{Far-UV to optical spectra of five Galactic O    
stars were analyzed: HD 216898 (O9IV/O8.5V), HD 326329 (O9V), HD 66788 (O8V/O9V),    
$\zeta$ Oph (O9.5Vnn), and HD 216532 (O8.5V((n))). We used a grid of TLUSTY    
models to obtain effective temperatures, gravities, rotational    
velocities, and to identify wind lines. The wind parameters of each object    
were obtained by using expanding atmosphere models from the CMFGEN code.}       
{We found that the spectra of our sample have mainly a photospheric    
origin. A weak wind signature is seen in \civ, from where    
mass-loss rates consistent with previous CMFGEN results    
regarding O8-O9V stars were obtained ($\sim 10^{-10}-10^{-9}$ M$_\odot$ yr$^{-1}$).    
A discrepancy of roughly 2 orders of magnitude is found between these    
mass-loss rates and the values predicted by theory ($\dot{M}_{Vink}$),    
 confirming a breakdown or a steepening of the modified wind   
momentum-luminosity relation at log $L_\star/L_\odot \lesssim   
5.2$. We have estimated the carbon abundance for the stars of our sample and      
concluded that its uncertainty cannot cause the {\it weak wind problem}.   
Upper limits on $\dot{M}$ were established for all objects using lines       
of different ions, namely, \pv, \ciii, \nv, \siiv, and \niv.    
All the values obtained are also in disagreement with theoretical    
predictions, bringing support to the reality of {\it weak winds}.    
Together with \civ, the use of \nv\, results in the lowest    
mass-loss rates: the upper limits indicate that $\dot{M}$ must be {{\it less}} 
than about -1.0 dex $\dot{M}_{Vink}$. Regarding the other    
transitions, the upper limits obtained still point to low rates:    
$\dot{M}$ must be {{\it less}} than about $(-0.5 \pm 0.2)$ dex     
$\dot{M}_{Vink}$. We studied the behavior of the H$\alpha$ line with different    
mass-loss rates. For two stars, only models with very low $\dot{M}$'s provide    
the best fit to the UV and optical spectra. We explored ways to fit the   
observed spectra with the predicted mass-loss rates ($\dot{M}_{Vink}$).    
By using large amounts of X-rays, we verified that few wind emissions takes    
place, as in {\it weak winds}. However, unrealistic X-rays luminosities had    
to be used (log $L_X/L_{Bol} \gtrsim -3.5$). The validity of the models    
used in our analyses is discussed.}         
{}             
    
        
\authorrunning{Marcolino et al.}        
\titlerunning{Analysis of late-type O dwarfs}        
        
\maketitle        
     
        
\section{Introduction:}        
      
Massive stars of spectral types O and B play an extremely important       
role in astrophysics. They possess high effective temperatures       
($T_{eff} >10$kK), intense ionizing radiation fields and often strong       
mass-losses through stellar winds, making their description considerably difficult       
for atmosphere and stellar evolution models. These stars are        
known to be progenitors of fascinating objects such as Red Supergiants (RSGs),       
Luminous Blue Variables (LBVs), Wolf-Rayet stars (W-Rs), and thus       
also of some of the most energetic phenomena in the Universe, i.e.,       
of type II supernovae and some $\gamma$-ray bursts (Massey 2003;       
Woosley \& Bloom 2006). They also heavily affect their       
host galaxies by transferring momentum, energy and enriched       
chemical elements to the interstellar medium (Abbott 1982; Freyer et al. 2003).        
        
Although they have been studied for decades, the properties, origin         
and evolution of OB stars still present several observational and         
theoretical challenges. The dependency of their mass-loss rates ($\dot{M}$)        
on the metallicity ($Z$) for example, as well as their effective temperatures        
and wind structure (e.g. clumping) have been continuously debated in the literature         
during the last years (see for example Vink et al. 2001;        
Martins et al. 2002; Bouret et al. 2005; Puls et al. 2006; Crowther et al. 2006).         
         
Among several interesting issues currently under discussion        
(for a review see Puls 2008; Hillier 2008), one that        
have been receiving special attention is the so-called {\it weak wind problem},        
which is posed by late-type O dwarf stars. From a qualitative point of view,        
O stars with weak winds present mainly an absorption spectrum, with the exception        
being a very few weak wind lines. In fact, often only a weak \civ\,  in P-Cygni is seen.        
In contrast, mid- and early O dwarfs can present P-Cygni profiles        
in lines such as \civ, \niv, \nv\, or \ov\, (see e.g. Snow et al. 1994; Escolano et al. 2008).       
Quantitatively, weak wind stars are defined as having mass-loss rates of less than about        
$10^{-8}$ M$_\odot$ yr$^{-1}$. Bouret et al. (2003) were one of the first to       
suggest such low values after analyzing O dwarfs in the H II region NGC 346 in       
the SMC. The spectra of three objects of their sample could only be reproduced       
by models using mass-loss rates of $\sim 10^{-10}$ to $10^{-9}$ M$_\odot$      
yr$^{-1}$. Similar results were found by Martins et al. (2004) for 4 O dwarfs       
in the compact star formation region N81, also in the SMC. Later, weak winds       
were also found in some Galactic O dwarfs, demonstrating that they       
are not result of an environmental (i.e. $Z$) effect (Martins et al. 2005).      
Despite these findings, as we will discuss later, very low       
$\dot{M}$ values are not of general consensus.       
    
There are interesting questions that O stars with very low mass loss       
rates rise. From the stellar evolution point of view, it is well       
known that mass-loss is a fundamental ingredient in the models.       
Very different values for this parameter can alter considerably       
the way the stars evolve. For example, mass-loss can change       
the rotational structure of a star (decreasing surface       
and internal velocities; $\Omega _r$) due to removal of angular       
momentum combined with internal transport mechanisms       
(Meynet \& Maeder 2000). Very low $\dot{M}$'s might thus imply       
that stars can keep high rotational velocities and get closer       
to break-up velocities. Low mass-loss rates can also change the       
way we understand the evolution in the LBV and W-R phases.       
The total mass lost from the Main Sequence prior the LBV       
phase can be much less than currently thought. As a consequence,       
in order to be consistent with observed masses of hydrogen deficient       
W-R stars, other intense mass-loss mechanisms must occur (e.g.       
continuum driven giant eruptions in the LBV phase) and/or       
evolutionary time scales (e.g. of the WNL phase) must be changed       
(see Smith \& Owocki 2006; van Marle et al. 2008). Given these        
considerations, to be sure that weak winds exist is now       
essential.      
 
From the radiative wind theory point of view, the low mass-loss rates       
obtained for late type O dwarfs also present challenges. While       
the last, state-of-art theoretical $\dot{M}$ predictions of Vink et al.       
(2000; 2001; $\dot{M}_{Vink}$) present a good match for objects with log $L/L_\odot$ greater       
than about 5.2 (neglecting clumping), for late-type, less luminous objects a discrepancy       
of even a factor of 100 can be found (Martins et al. 2005). Late O dwarfs data    
also suggest a breakdown of the modified wind momentum luminosity relation    
(WLR; Kudrizki \& Puls 2000) or at least a change in its slope (see Fig. 41 of Martins et al. 2005).      
All these facts constitute what is now called the {\it weak wind problem}    
\footnote{Another type of weak wind problem exists, which is the weaker      
wind signatures of some stars compared to others of the same spectral type     
(e.g. $\theta^1$ Ori C; Walborn \& Panek 1984). Throughout this paper, we consider     
"weaker winds" compared to the predicted mass-loss rates according to     
Vink et al. (2000) and to normal O stars of earlier spectral types.}.   
       
A possible reason for the discrepancies aforementioned        
is that mass-loss rates determinations based on UV lines       
could be incorrect, as it was suggested by Mokiem et al. (2007).      
The use of only one UV diagnostic line (i.e. \civ), the abundance of       
carbon, the effect of X-rays, and the wind ionisation structure       
derived from the models are claimed to be considerable        
sources of errors for the derivation of $\dot{M}$.       
This means that stronger winds could perhaps have been found if appropriate        
diagnostic tools were used/available. According to Mokiem et al. (2007), three        
objects which are supposed to be in the weak wind regime ($\zeta$ Oph,        
CygOB2\#2, and HD 217086) actually do not show very low mass-loss        
rates (i.e. $\dot{M}$'s are consistent with theory) if H$\alpha$ is used        
as the diagnostic.       
      
In order to clarify some of the issues described above and get        
more insight into the weak wind problem, we have analyzed in detail     
far-UV, UV and optical high-resolution spectra of five Galactic late-type  
O dwarf stars. Their stellar and wind physical parameters were obtained using 
the codes TLUSTY and CMFGEN (Hubeny \& Lanz 1995; Hillier \& Miller 1998). 
With this work, the number of O8-9V objects studied by means of  
state-of-art atmosphere models is increased considerably, since only  
six were previously analyzed (Repolust et al. 2004; Mokiem et al. 2005;  
Martins et al. 2005).
 
The remainder of this paper is organized as follows. In Section
\ref{observations} we describe the way we have selected our targets and the     
observational material used. A description of the atmosphere codes and the adopted assumptions    
are given in Section \ref{modelsass}. The analysis of each object of our sample     
is presented in Section \ref{analysis}. The derived stellar and wind
parameters are presented later in Section \ref{summary} along with comparisons 
to previous results and theoretical predictions. In Section \ref{carbona} 
we discuss the carbon abundance and its relation with the mass-loss 
rate. In Section \ref{mdotlateo} we present estimates for the mass-loss rates 
of the programme stars using other lines besides \civ. In Section
\ref{discussion} we discuss the consequences of our findings and explore a    
way to have agreement with the observed spectra using the mass-loss rates 
predicted by theory. Section \ref{conclusions} summarizes the main results found in our study.   
    

\section{Target Selection and Observations:}        
\label{observations}

\begin{table*}[]
\centering \caption{Observational Data of the Programme Stars.}
\label{tab_obs}
\begin{tabular}{lcccccccc}
\hline
\hline
        &                &          & \multicolumn{2}{c}{far-UV} & \multicolumn{2}{c}{UV} & \multicolumn{2}{c}{Optical}  \\
Star & Spectral Type$^{\mathrm{a}}$ &  V & Data Set$^{b}$ & Date & Data Set & Date & Instrument & Date  \\
\hline
  HD 216898  & O9IV, O8.5V & 8.04 & F - A0510303 & 2000-08-03 & SWP43934 & 1992-02-05 & OHP/ELODIE & 2004-08-29      \\
  HD 326329  & O9V         & 8.76 & F - B0250501 & 2001-08-09 & SWP48698 & 1993-09-21 & ESO/FEROS  & 2005-06-25      \\
  HD 66788   & O8V, O9V    & 9.43 & F - P1011801 & 2000-04-06 & SWP49080 & 1993-11-03 & ESO/FEROS  & 2004-12-22 \\
 $\zeta$ Oph & O9.5Vnn     & 2.56 & C - C002     & 1972-08-29$^{\mathrm{\dagger}}$ & SWP36162 & 1989-04-29 & OHP/ELODIE & 1998-06-11  \\
  HD 216532  & O8.5V((n))  & 8.03 & F - A0510202 & 2000-08-02 & SWP34226 & 1988-09-11 & OHP/ELODIE & 2005-11-10 \\
  \hline
\end{tabular}
\label{tabone}
  \begin{list}{}{}
\item[$^{\mathrm{a}}$] Spectral types come from Lesh (1968), Schild et al. (1969), Garrison (1970), Walborn (1973), and MacConnell \& Bidelman (1976).
\item[$^{\mathrm{b}}$] Satellite used to obtain spectra: F=FUSE and C=Copernicus.
\item[$^{\mathrm{\dagger}}$] Orbits \#115-247 (Snow et al. 1977).
\end{list}
\end{table*}          
    
The Galactic stars claimed to have weak winds in the literature belong     
mainly to the O8, O8.5, O9, and O9.5V spectral types (hereinafter we refer     
to them simply as O8-9V). In order to carry out our investigation, we     
initially selected a sample of several objects belonging     
to these classes by using the Galactic O Star Catalog (Ma\`iz-Apell\'aniz     
et al. 2004; hereinafter the GOS Catalog). Various stars were previously observed     
by our group (at optical wavelengths), and we have retrieved the spectra of     
some others from public available archives. After a first examination, we neglected     
all those objects that presented a peculiar spectrum. This included for     
example stars known to be binaries (having contamination or known     
to show strong variability in the spectrum) or ON8-9V objects.     
Due to our aim to study wind lines, we have drastically decreased     
the number of objects by considering only the ones having both UV and     
far-UV data. Furthermore, we have also neglected the stars in     
common with the work of Martins et al. (2005), since they were analyzed    
with the same atmosphere code utilized here (CMFGEN). In the end, we have      
chosen a subsample of five objects, listed in Table \ref{tab_obs}.
   
\subsection{Optical Data:}    
      
For the optical, we used data collected with the MPI 2.2m telescope,     
located at the European Southern Observatory (ESO), in La Silla, Chile.    
The high resolution (R=48000) FEROS spectrograph was     
used and the total wavelength coverage is $\sim$ 4000-9000\AA.     
The spectra were wavelength calibrated and optimally extracted     
using the available pipeline (for more details see Kaufer et al. 1999).     
    
We have also used spectra obtained with the 1.93m telescope at the     
Observatoire de Haute-Provence, France, using ELODIE.     
This spectrograph has a resolving power R = 42000 in the wavelength range      
$\sim$3900-6800\AA. The exposure times were set to yield a     
signal-to-noise (S/N) ratio above 100 at about 5000\AA.     
The data reduction, the order localization, background     
estimate/subtraction, and wavelength calibration, were performed     
using the available pipeline (see Baranne et al. 1996).    
Each order was normalized by a polynomial fit to the continuum,     
specified by carefully selected continuum windows. At last, we have     
merged the successive orders to reconstruct the full spectrum.     
The spectrum of $\zeta$ Oph was obtained from the      
ELODIE archive (see Moultaka et al. 2004 for more details).    
We have retrieved small portions of its pipeline treated spectrum     
and then normalized them using polynomial fits.    
    
\subsection{Ultraviolet Data:}    
    
Different sources were used for the UV and far-UV data. We have retrieved      
spectra from the IUE and FUSE satellites using the Multimission     
Archive at STScI\footnote{MAST - http://archive.stsci.edu/}.      
Regarding the IUE data, the high resolution SWP mode was       
preferred ($\sim 0.2$\AA), but spectra in the LWP/LWR region     
were also used to help in the determination of reddening     
parameters (R and E(B-V)). Regarding the FUSE data, we have     
relied only on the LiF2A channel, which comprises the $\sim 1087$-$1182$\AA\,     
interval. Spectra of a same object were co-added and smoothed with     
a five point average for better signal-to-noise ratio (S/N) and clarity.     
Although small, the LiF2A region provided two very useful features     
for our study, namely, \pv\, and  \ciii\, (see Section \ref{othermdot}).     
We have avoided other FUSE channels covering wavelengths less than     
$\sim 1087$\AA\, due to severe interstellar contamination.     
For one object of our sample, $\zeta$ Oph, there is no FUSE data     
available. Therefore, we have used far-UV data obtained with    
the {\it Copernicus} (``OAO-3'')  satellite. The spectrum was retrieved    
from the Vizier Service\footnote{http://webviz.u-strasbg.fr/viz-bin/VizieR}    
and normalized ``by eye''.    
   
        
\section{Models and Assumptions:}        
\label{modelsass}
 
In order to obtain the stellar and wind parameters of the stars of     
our sample we have used the well known atmosphere codes        
TLUSTY (Hubeny \& Lanz 1995) and CMFGEN (Hillier \& Miller 1998).        
The TLUSTY code adopts a plane-parallel geometry, assumes hydrostatic and       
radiative equilibrium, and a non-LTE treatment including line-blanketing is taken into        
account. As such, it is suitable only to model lines that        
are not formed in a stellar wind, i.e., of photospheric origin.        
On the other hand, the radiative transfer and statistical equilibrium        
equations in CMFGEN are solved in a spherically symmetric outflow.        
The effect of line-blanketing is also taken into account, via a        
super-levels formalism (for more details see Hillier \& Miller 1998).        
For the outflow we assume a $\beta$ velocity law which connects         
smoothly (at depth) with a hydrostatic structure provided by TLUSTY.           
 
\subsection{Methodology:}      
\label{method}      
    
To start our analysis, we used a new grid of TLUSTY model atmospheres based    
on the OSTAR2002 grid (Lanz \& Hubeny 2003). The new grid uses finer sampling   
steps in effective temperature ($1000K$) and surface gravity (0.2 dex),    
updated model atoms (for Ne and S ions, \ion{C}{iii}, \ion{N}{ii}, \ion{N}{iv},   
\ion{O}{ii}, and \ion{O}{iii}), and additional ions (\ion{Mg}{ii},   
\ion{Al}{ii}, \ion{Al}{iii}, \ion{Si}{ii}, and \ion{Fe}{ii}; see Lanz \&   
Hubeny 2007). In order to derive $T_{eff}$, we have used    
mainly \hei\, and \heii\, as diagnostic lines, but fits to other    
transitions such as \heic\, and \heiic\, were also checked for consistency.        
Typical errors for $T_{eff}$ range from 1 to $2kK$. Our derivation of log $g$   
was based on the fit to the H$\gamma$ wings. For this parameter, the   
uncertainty varies from 0.1 to 0.2 dex, depending on the object. The   
rotational velocities ($vsini$) were adopted from previous studies (e.g.    
Penny 1996; Howarth et al. 1997) and/or refined/estimated from the fitting    
process when necessary. After we have derived the basic photospheric   
properties (i.e. $vsini$, log $g$, and $T_{eff}$) we have switched to CMFGEN    
to continue our investigation. As in previous studies, we noted a good   
agreement between TLUSTY and CMFGEN (e.g. Bouret et al. 2003). In general, conspicuous discrepancies   
were sorted out when we have increased the number of species or atomic        
levels and/or super-levels in CMFGEN.       
   
Once effective temperatures were obtained, we have computed the radii by    
adopting luminosities typical of the O8-O9 dwarfs (see Table 4 of Martins    
et al. 2002), following the equation $R_\star=(L_\star/4\pi \sigma   
T^4_{eff})^{1/2}$. As we did not normalize the UV spectra, we    
have used the reddening parameters R and E(B-V) (following Cardelli et al. 1988)    
and the distance as free parameters to match the continuum in the IUE region.  
When a good fit was achieved, the R, E(B-V), and the distance used were  
considered representatives for the object in question. The values obtained were    
generally consistent with the ones estimated in the literature (see       
Sect. \ref{analysis}). Whenever a large discrepancy was encountered,        
we have revised our adopted $L_\star$ and hence $R_\star$,        
keeping $T_{eff}$ constant. A reasonable agreement is found        
between the final models and the observed absolute fluxes in        
the UV. In some cases however, local scalings of the        
theoretical continuum had to be made.       
    
The two main wind parameters to be obtained are the       
mass-loss rate and the terminal velocity. Ideally, the       
determination of the mass-loss rates should       
be done by fitting different (sensitive) P-Cygni and       
emission lines in the optical (e.g. H$\alpha$) and in       
the UV (e.g. \civ, \nv, \siiv). In the case of       
late-type O dwarfs however, most of these wind transitions       
are absent and the best estimator of $\dot{M}$ is \civ.     
Regarding the velocity structure, we have assumed $\beta = 1$ for      
all our models. Initial tests with $\beta$ ranging from       
0.8 to 1.6 did not improve the quality of the \civ\, fits.      
The terminal velocity ($v_\infty$) is a very difficult       
parameter to obtain for O8-O9V stars, since there are no       
clear, saturated P-Cygni profiles. We have chosen to       
first use previous $v_\infty$ estimates in the literature       
and then change its value from the fits, if needed.       
When such estimates were not available, we have started our       
models assuming $v_\infty$ values between 1300-1700 km s$^{-1}$.      
We conservatively assume the uncertainty in this parameter       
to be about 500 km s$^{-1}$. Regarding the mass-loss rate, we    
consider a conservative uncertainty of $\sim$ 0.7 dex, following    
the analysis performed by Martins et al. (2005).
     
\begin{table}[]    
\caption{Atomic Data Used in the CMFGEN Models.}    
\label{atomic}    
\begin{tabular}{l|cccc}    
\hline    
\hline    
   Ion      &  \multicolumn{2}{c}{Basic Models} & \multicolumn{2}{c}{Full Models}   \\     
            &  $\#$ levels &  $\#$ super-levels &   $\#$ levels &  $\#$ super levels \\    
\hline    
H~{\sc i}   &        30    &     30   &     30       &  30     \\    
He~{\sc i}  &        69    &     69   &     69       &  69     \\    
He~{\sc ii} &        30    &     30   &     30       &  30     \\    
C~{\sc ii}  &        22    &     22   &    338       & 104     \\    
C~{\sc iii} &       243    &     99   &    243       & 217     \\    
C~{\sc iv}  &        64    &     64   &     64       &  64     \\    
N~{\sc iii} &       287    &     57   &    287       &  57     \\    
N~{\sc iv}  &        70    &     44   &    70        &  44     \\    
N~{\sc v}   &        49    &     41   &    49        &  41     \\    
O~{\sc ii}  &       296    &     53   &   340        & 137     \\    
O~{\sc iii} &       115    &     79   &   115        &  79     \\    
O~{\sc iv}  &        72    &     53   &    72        &  53     \\    
Ne~{\sc ii} &        -     &      -   &   242        &  42     \\    
Mg~{\sc ii} &        -     &      -   &    65        &  22     \\    
Al~{\sc iii}&        -     &      -   &    65        &  21     \\    
Si~{\sc iii}&        -     &      -   &    45        &  25     \\    
Si~{\sc iv} &        33    &     22   &    33        &  22     \\    
P~{\sc v}   &        62    &     16   &    62        &  16     \\    
S~{\sc iv}  &        -     &      -   &    142       &  51     \\    
S~{\sc v}   &        -     &      -   &    216       &  33     \\    
Fe~{\sc iii}&        -     &      -   &   607        &  65     \\    
Fe~{\sc iv} &      1000    &     64   &  1000        &  64     \\    
Fe~{\sc v}  &      1000    &     45   &  1000        &  45     \\    
\hline    
\end{tabular}    
\end{table}     
    
We have assumed a fixed microturbulence $\xi _t$ of 15 km s$^{-1}$       
in the co-moving computation of our models. The CMFGEN code allows however       
a description of a depth dependent microturbulent velocity when       
passing to the observer's frame, following the equation $\xi (r) = \xi _{min} + (\xi _{max} -       
\xi _{min})v(r)/v_\infty$. We have fixed $\xi _{min}$ at 5 km s$^{-1}$       
(near the photosphere) and for $\xi _{max}$ we have assumed a value       
of $0.1 v_{\infty}$. As it was already reported in previous studies,       
we noted that slightly different values for these parameters       
do not present significant changes in the spectrum.       
    
We have used two sets of atomic data for CMFGEN in our study. The     
first fits/models to the observed spectra, tests, and the exploration     
of higher mass-loss rates (Section \ref{mdotlateo}) were done     
using {\it basic models}, i.e., models with a reduced number of species and     
levels/super-levels. For our final model for each object, a more complete
({\it full}) set was used. In total, 4.5GB (2GB) of memory space (RAM) was required     
to compute each {\it full} ({\it basic}) model. The details of the atomic data     
are shown in Table \ref{atomic}. This approach could save a lot of     
computational time and resources, without compromising the results.     
The main difference between these two sets of data is the quality of the     
optical fit. Some lines could only be reproduced after the inclusion     
of additional species/ions (e.g. Mg~{\sc ii} $\lambda 4481$ and     
Si~{\sc iii} $\lambda 4553$) or by increasing the number of     
levels and/or super-levels. In the ultraviolet, only the 1850-2000\AA\,     
region is affected by the heavier data set (being better fitted), due to the     
inclusion of Fe~{\sc iii}. In other parts the changes are minor.    
A solar chemical abundance was adopted for all elements (following    
Grevesse \& Sauval 1998). As it will be seen later in Section   
\ref{analysis}, it provides reasonable fits to the observations.    
In this paper, only the amount of carbon is investigated in    
detail (see Section \ref{carbona}), given its direct    
relation with the weak wind problem through \civ.   
   
\subsection{Clumping and X-rays:}      
\label{cxrays}   
      
Currently, the effect of wind density inhomogeneities (i.e. clumping)        
is implemented in CMFGEN through a depth dependent volume filling       
factor $f = f_\infty + (1 -  f_\infty)e^{v(r)/v_{cl}}$. The parameter       
$v_{cl}$ regulates the velocity where clumping starts to       
be important. At velocities higher than $v_{cl}$, $f$ converges       
quickly to $f_\infty$. With this description, the model is       
homogeneous near the photosphere. Because we have a lack of       
wind lines in the spectra of the stars studied here, we       
chose to not use clumping in our models, i.e., $f_\infty = 1$.      
We stress that the use of clumping usually requires a decrease       
in the mass-loss rate of a (previously homogeneous final) model by a factor       
of $1/\sqrt{f_\infty}$, in order to fit (again) the observed spectra.      
Thus, the mass-loss in our final models can perhaps be       
overestimated by a factor of about three if $f_\infty \sim 0.1$ in       
the wind of these objects. In any case, to not use clumping       
means that we might be biased towards stronger winds.      
      
CMFGEN also allows the inclusion of X-rays in the models to       
simulate the effects of shocks due to wind instabilities.      
The X-rays emissivity is basically controlled by a shock       
temperature plus a filling factor parameter and it is       
distributed throughout the atmosphere (see Hillier \& Miller 1998       
for details). The effects of X-rays in the atmosphere of       
O dwarfs were discussed in detail by Martins et al. (2005).       
These authors convincingly show that X-rays are not as important       
for early-type as they are for late-type O stars (see also Macfarlane et al. 1994).       
For early-type objects (denser winds), the ionisation and wind lines such 
as \civ\, do not present significant changes when X-rays are included.  
Only wind lines belonging to super-ions (\ovi\, and to a lesser extent, \nv) 
are affected. In contrast, for late-type stars it is observed that the ionisation       
structure and the \civ\, line changes considerably. Because the ionisation       
is shifted to C~{\sc v-vi}, mass-loss rates about ten times stronger       
than in a model without X-rays are sometimes required to fit the observed \civ\, line.       
Due to this fact, as it is the case of non-clumped models, higher mass-loss       
rates are favored when X-rays are included. In this paper, we use a fixed       
shock temperature of $3 \times 10^6 K$ and the filling factor is chosen        
to keep log $L_x/L_{Bol}$ close to -7.0 (within $\pm 0.1$). We recall       
that a log $L_x/L_{Bol}$ of about -7.0 is the typical value observed    
(see e.g. Sana et al. 2006). For HD 326329 however, we used models with a higher    
ratio, in accord with the observations (log $L_X/L_{Bol} \sim -6.5$;    
Sana et  al. 2006). For $\zeta$ Oph, we use a log $L_X/L_{Bol} \sim -7.3$    
(Oskinova 2005; Oskinova et al. 2006).   
    
\section{Spectral Analysis:}        
\label{analysis}       
           
In this section we present the analysis of our sample.    
A brief introduction about each object is made, followed     
by the presentation of our CMFGEN model fits to the UV and optical     
observed spectra. The discrepancies found are discussed and     
a comparison to previous results in the literature is given     
when necessary. 
   
\begin{figure*}        
\centering        
\includegraphics[width=14cm,height=6.1cm]{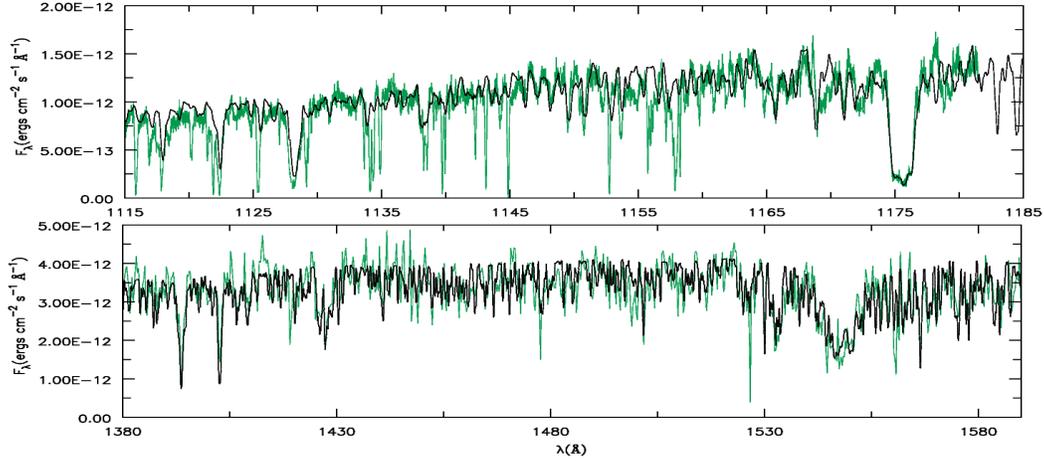}   
\caption{Ultraviolet spectra of HD 216898 (green/light gray line) and our final    
model (black line; log $\dot{M} = -9.35$).}      
\label{hd216a}        
\end{figure*}        
       
\begin{figure*}        
\centering       
\includegraphics[width=14cm,height=14cm]{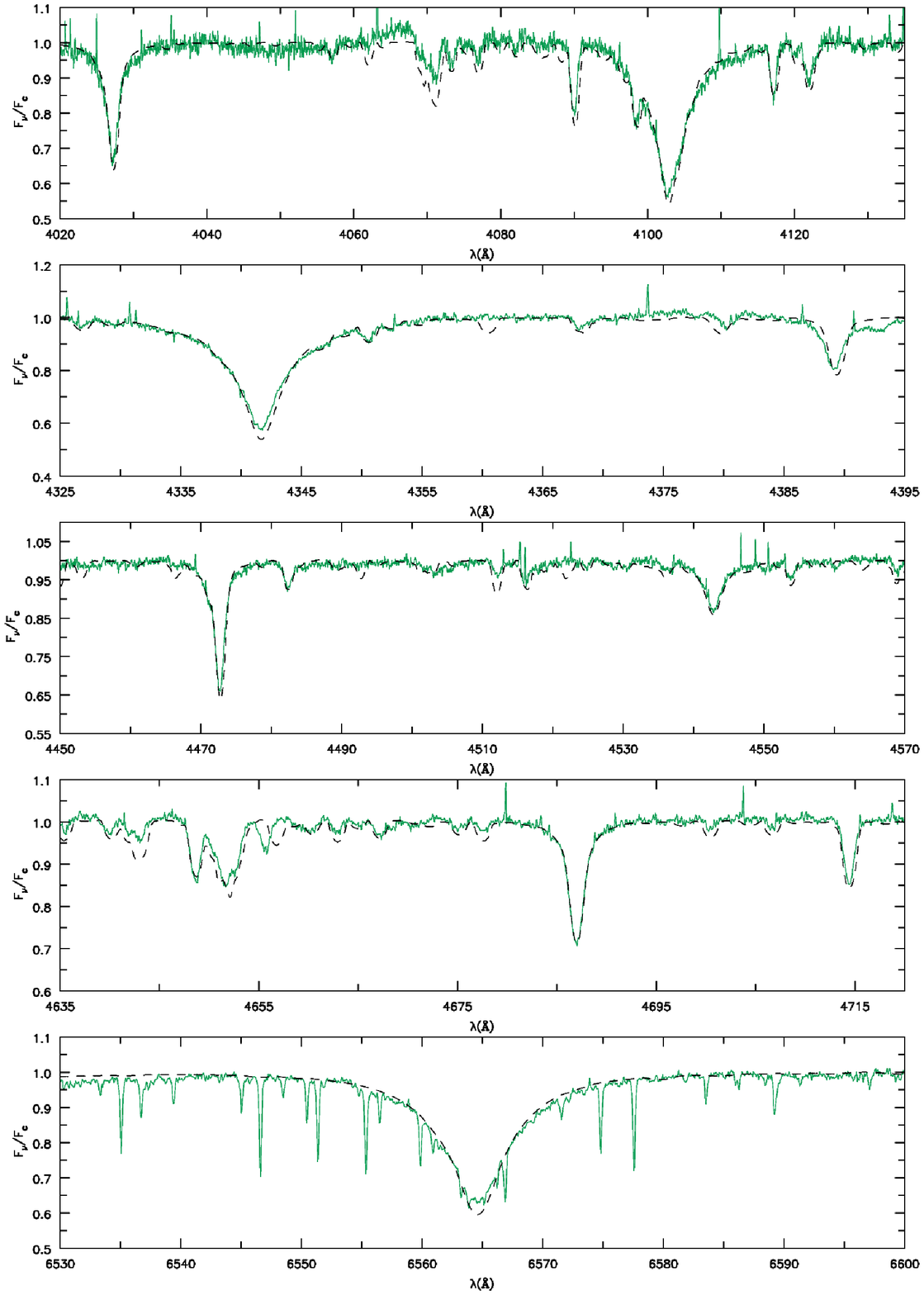}    
\caption{Optical spectrum of HD 216898 (green/light gray line) and our final model    
(black/dashed line).}        
\label{hd216b}        
\end{figure*}        
   
\begin{figure*}        
\centering        
\includegraphics[width=14cm,height=6.1cm]{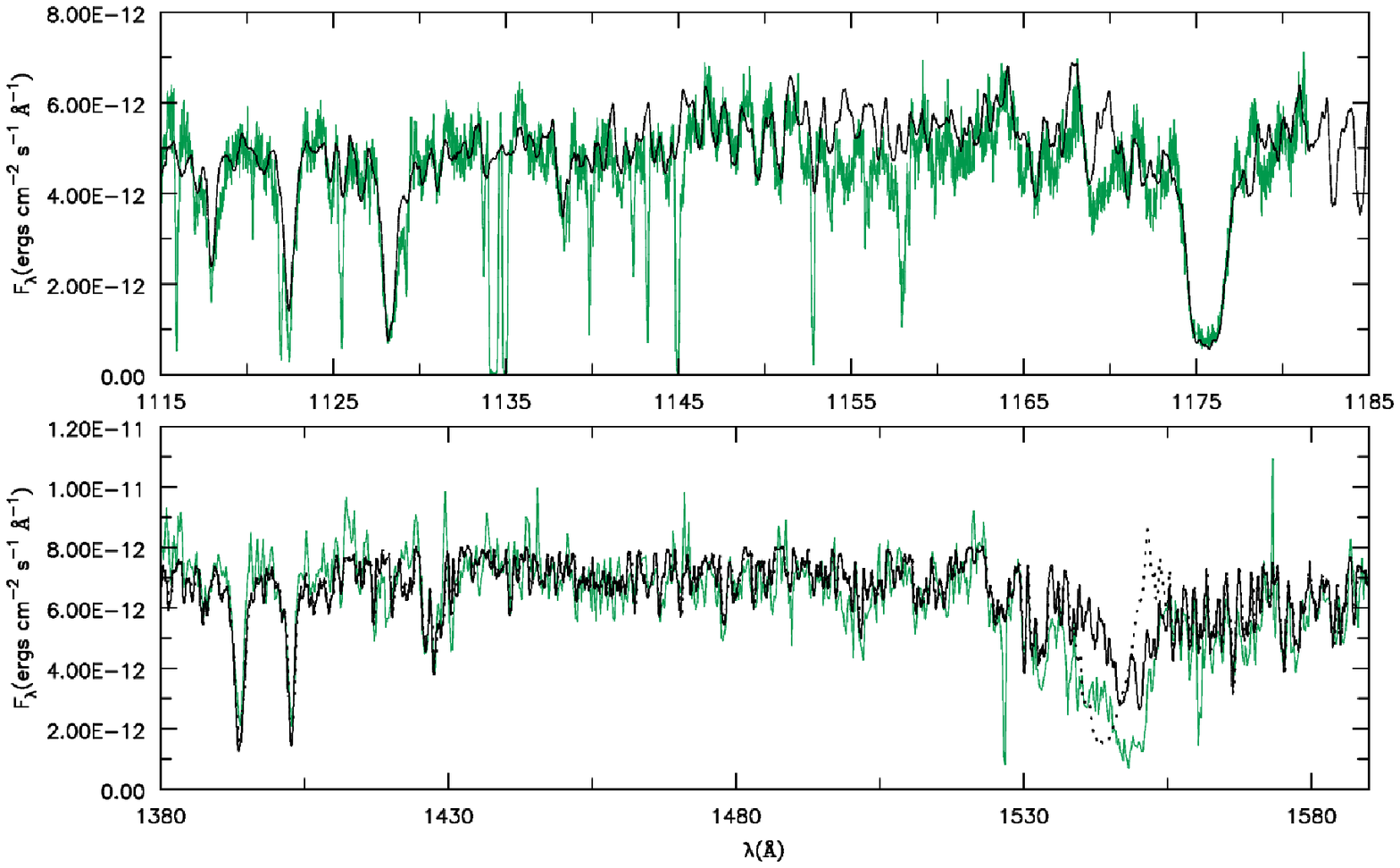}        
\caption{Ultraviolet spectra of HD 326329 (green/light gray line) and our final    
model (black line; log $\dot{M} = -9.22$). A model with a higher mass-loss rate is shown as a dotted    
line (log $\dot{M} = -8.55$; see text for more details).}        
\label{hd326a}        
\end{figure*}

\begin{figure*}        
\centering        
\includegraphics[width=14cm,height=14cm]{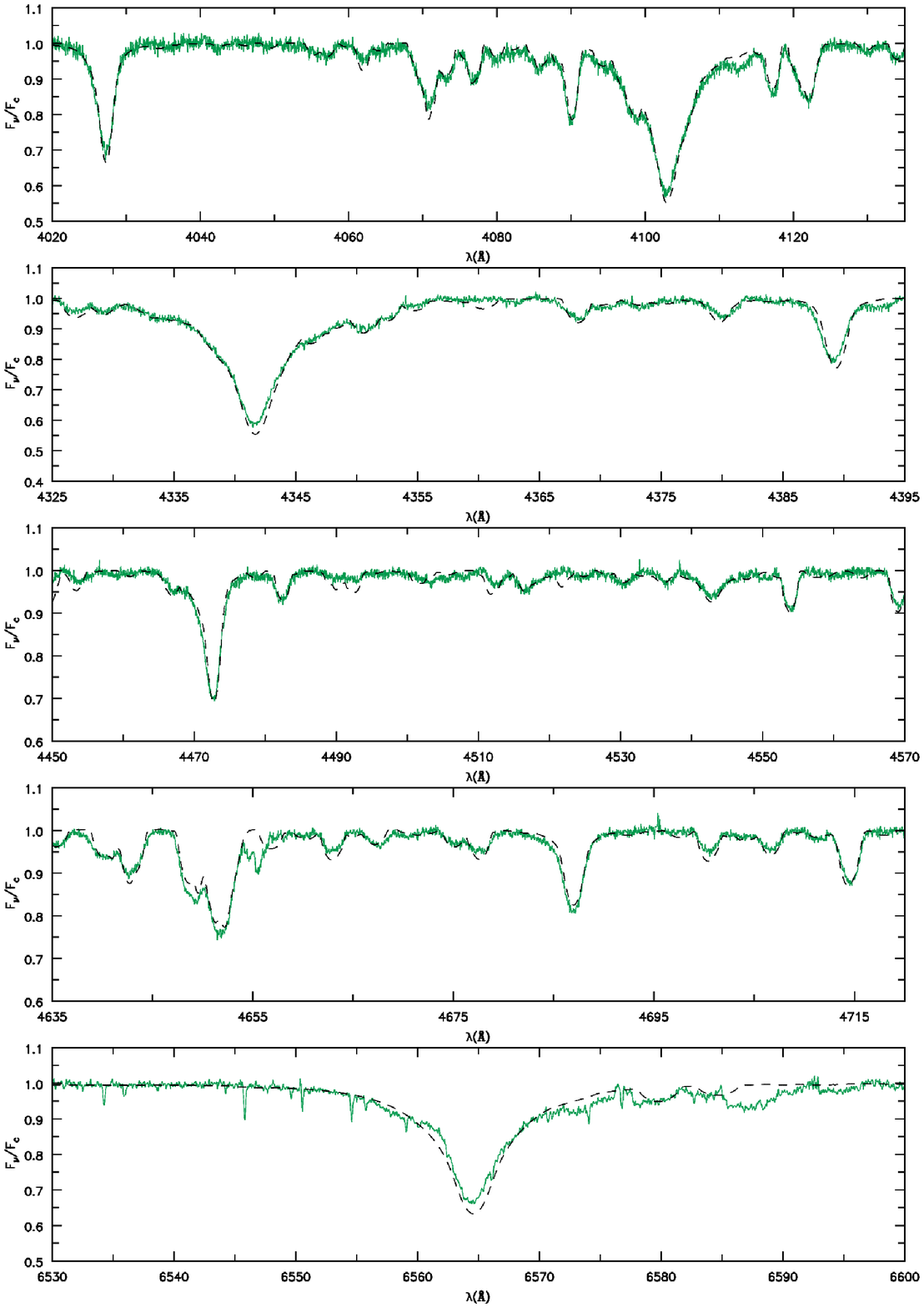}        
\caption{Optical spectrum of HD 326329 (green/light gray line) and our final model    
(black/dashed line).}        
\label{hd326b}        
\end{figure*}        
      
\begin{figure*}        
\centering       
\includegraphics[width=14cm,height=6.1cm]{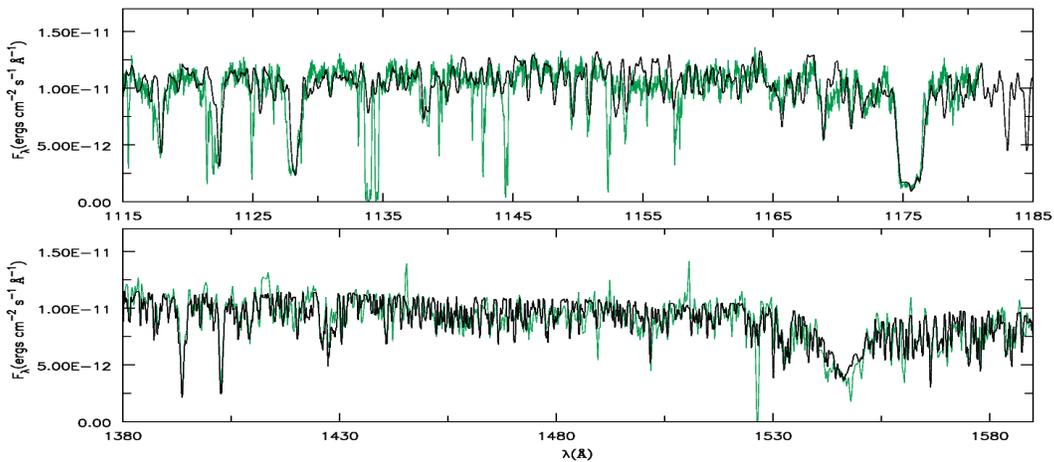}        
\caption{Ultraviolet spectra of HD 66788 (green/light gray line) and our final    
model (black line; log $\dot{M} = -8.92$).}        
\label{hd66a}        
\end{figure*}             
      
\begin{figure*}        
\centering       
\includegraphics[width=14cm,height=14cm]{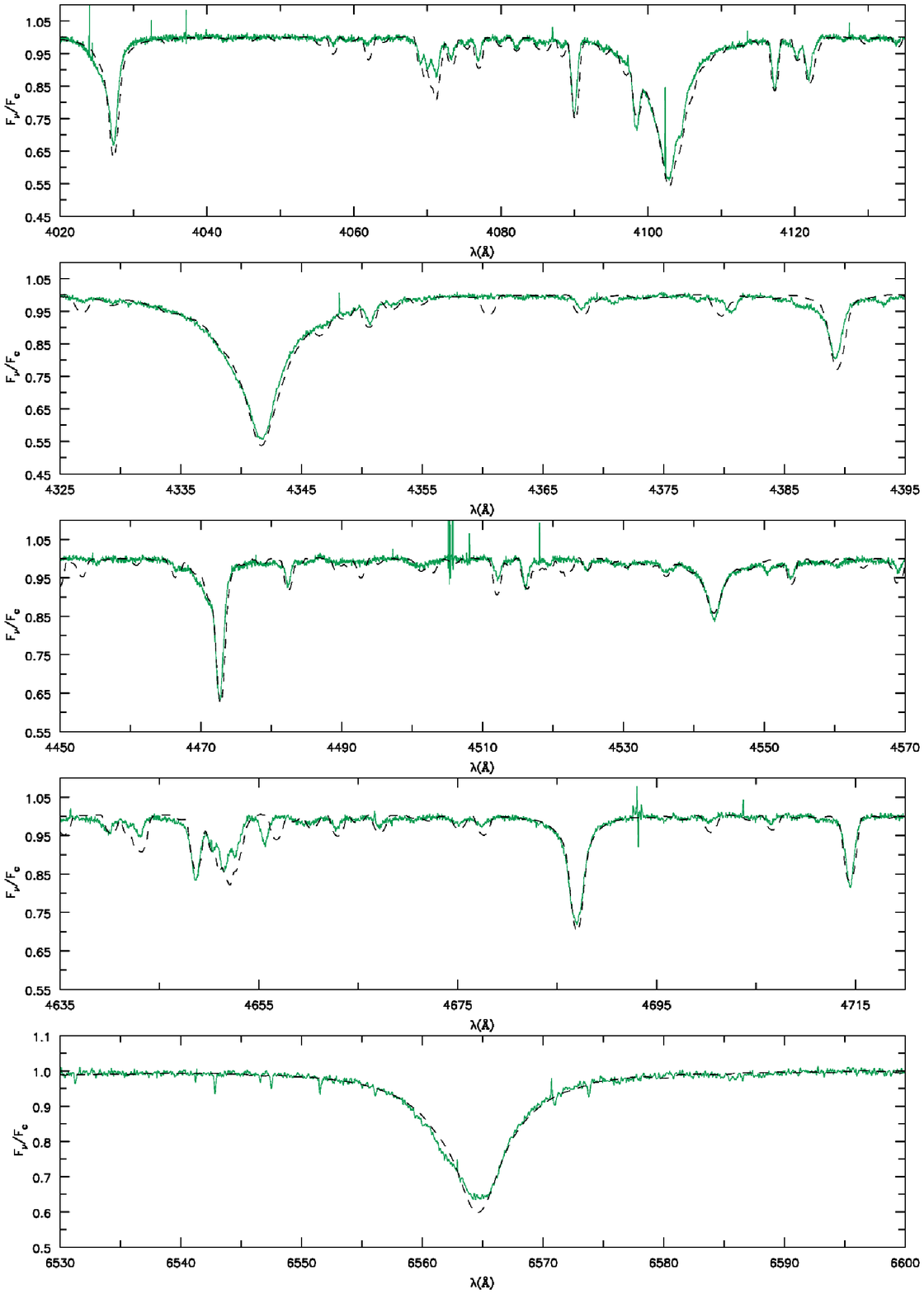}        
\caption{Optical spectrum of HD 66788 (green/light gray line) and our final model    
(black/dashed line).}        
\label{hd66b}        
\end{figure*}        
     
\begin{figure*}        
\centering        
\includegraphics[width=14cm,height=6.1cm]{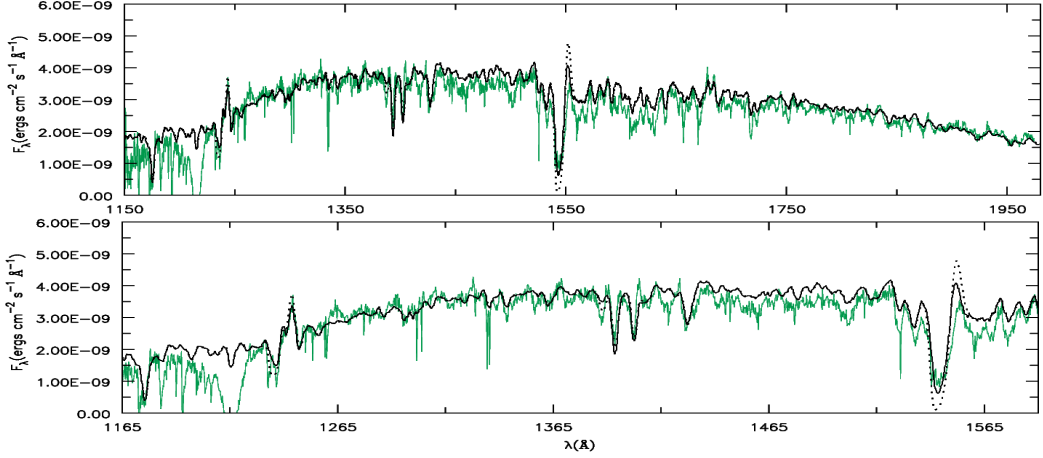}        
\caption{Ultraviolet spectrum of $\zeta$ Oph (green/light gray line) and our    
final model (black line; log $\dot{M} = -8.80$).   
A model with a higher mass-loss rate is shown as a dotted    
line (log $\dot{M} = -8.30$; see text for more details).}        
\label{zeta1}        
\end{figure*}           
        
\begin{figure*}        
\centering       
\includegraphics[width=14cm,height=14cm]{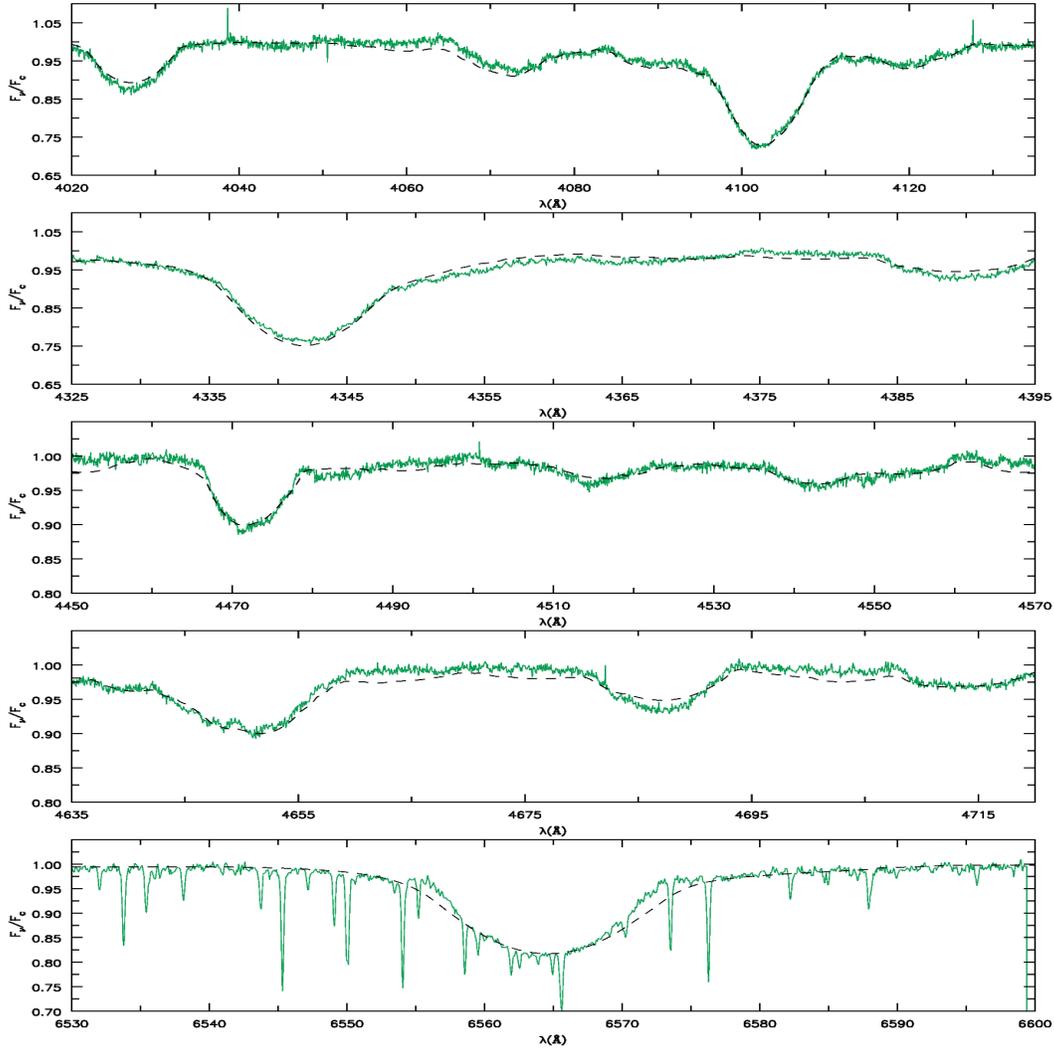}        
\caption{Optical spectrum of $\zeta$ Oph (green/light gray line) and our final   
  model (black/dashed line).}        
\label{zeta2}        
\end{figure*}        
         
\begin{figure*}        
\centering       
\includegraphics[width=14cm,height=6.1cm]{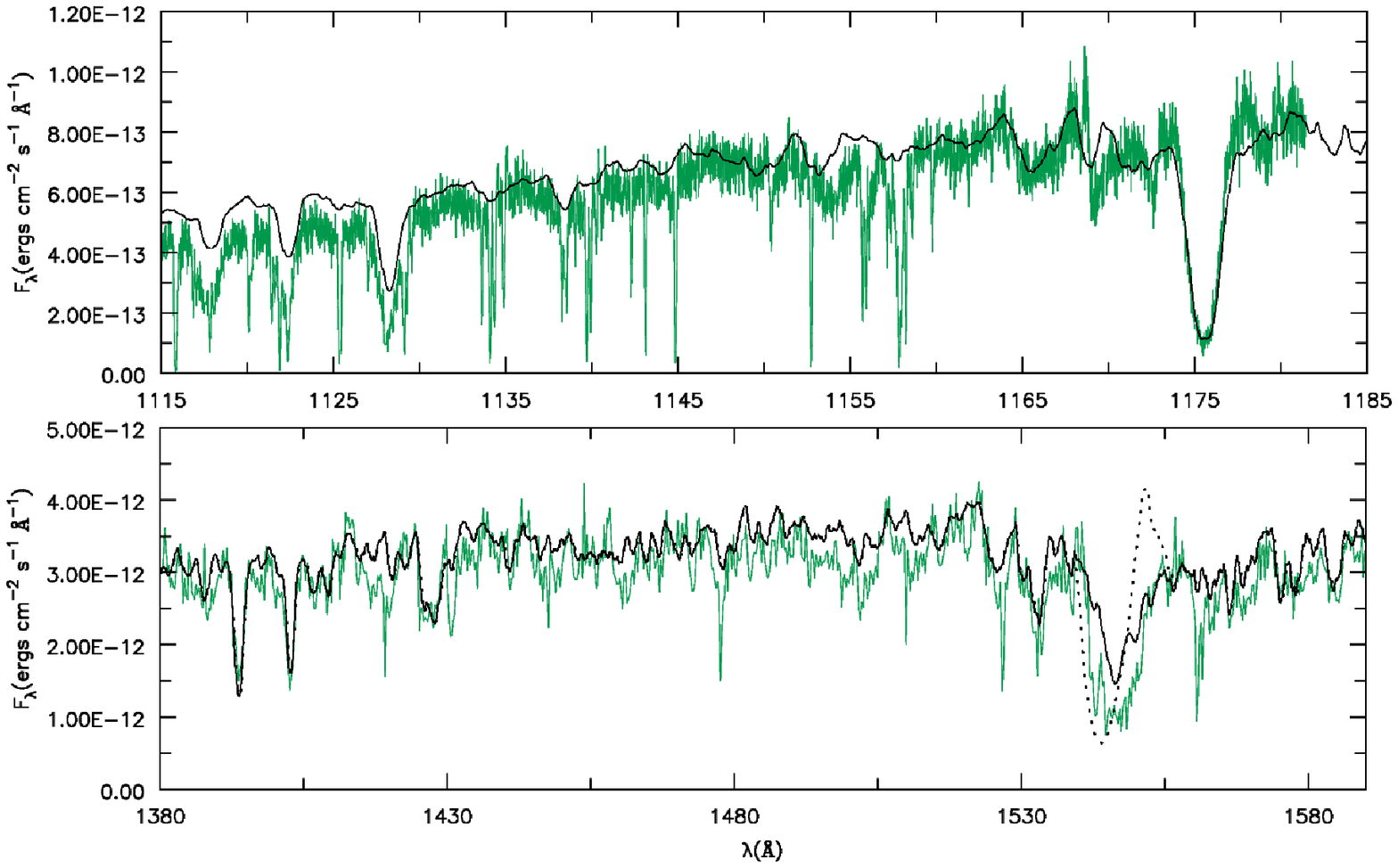}          
\caption{Ultraviolet spectra of HD 216532 (green/light gray line) and our final    
model (black line; log $\dot{M} = -9.22$). A model with a higher mass-loss rate is shown as a dotted    
line (log $\dot{M} = -8.78$; see text for more details).}        
\label{hd216532a}        
\end{figure*}        
      
\begin{figure*}        
\centering       
\includegraphics[width=14cm,height=14cm]{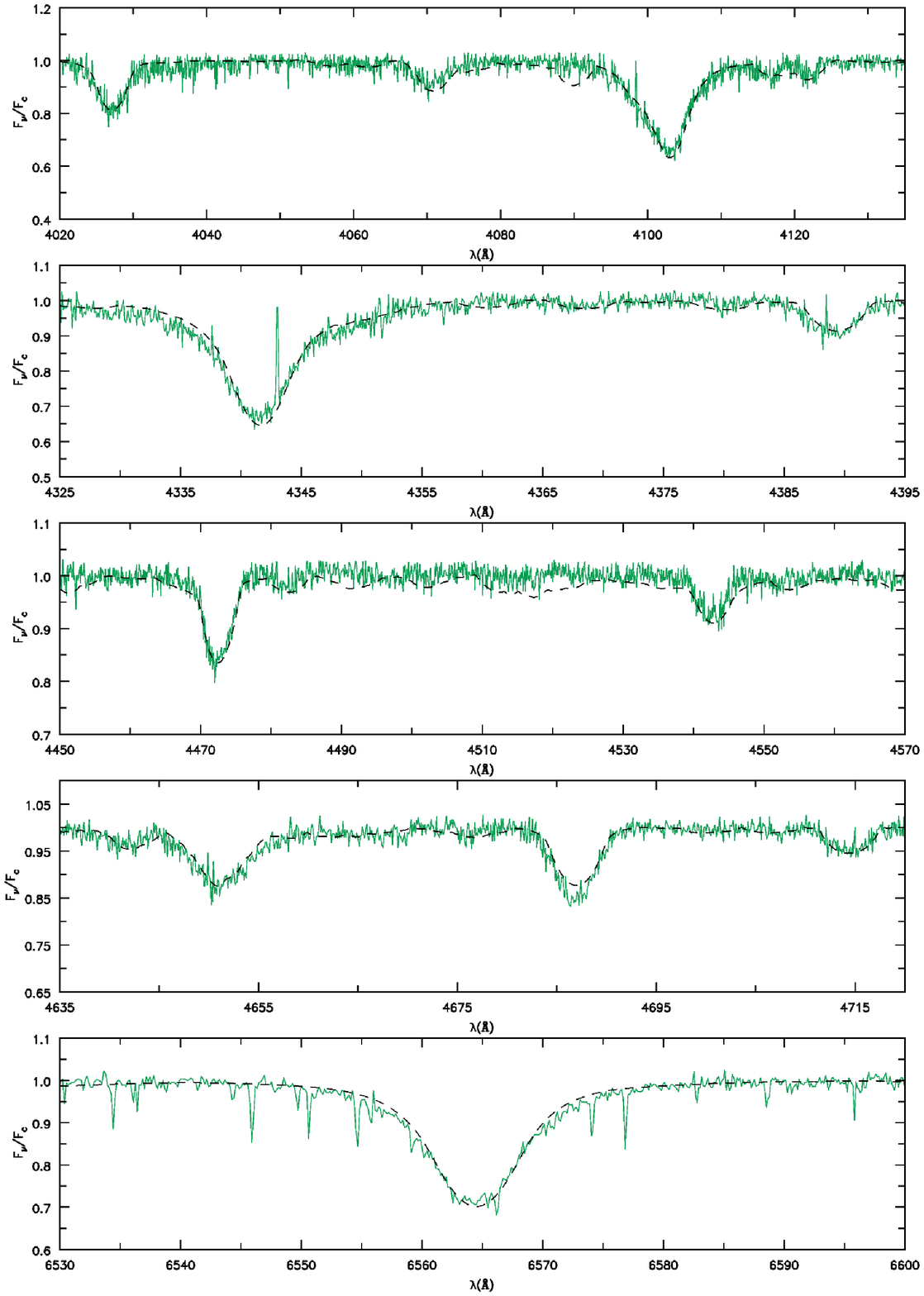}        
\caption{Optical spectrum of HD 216532 (green/light gray line) and our final    
model (black/dashed line).}        
\label{hd216532b}        
\end{figure*}        

\subsection{HD 216898}        
\label{hd216sec}     
      
HD 216898 belongs to the Cepheus OB3 association (Garmany \& Stencel 1992).
So far, no recent atmosphere models (i.e. unified, line-blanketed) were used
to analyze this object. Its mass-loss rate was investigated two decades ago by       
Leitherer (1988), who could only establish an upper limit     
(log $\dot{M} < -6.97$) from the H$\alpha$ profile.     
    
Our fits to the far-UV and UV spectra are presented in Figure \ref{hd216a}.
The E(B-V) and distance used are 0.7 and 1.1kpc, respectively.  
Four intense transitions can be noted in the FUSE region: \pv,  Si~{\sc iv} $    
\lambda\lambda 1122,28$, and \ciii. The various narrow deep absorptions that are not    
reproduced by our models are of  interstellar origin. As it can be seen in     
Figure \ref{hd216a} (bottom), our final model also fits correctly the main     
features of the IUE spectrum. In the 1380-1590\AA\, interval, \siiv\, and    
\civ\, are the most evident transitions. The other several       
absorptions observed we identified to be mainly due to Fe~{\sc iv} lines ({\it an iron       
forest}). We note that a TLUSTY model with the same photospheric parameters used in CMFGEN present a       
similar fit to the far-UV and UV, with the exception of \civ. In TLUSTY, this doublet is seen as       
two well separated absorptions, while the observations show an extended blue        
feature that is only reproduced with CMFGEN, i.e., it is formed in the wind.      
      
The mass-loss needed to fit \civ\, is log $\dot{M} = -9.35$. If we take the other physical       
parameters derived for this object ($T_{eff}$, $L_\star$, and $M_\star = gR_{\star}^2/G$)       
and use the recipe of Vink et al. (2000), we derive a predicted mass-loss rate
of log $\dot{M}_{Vink} = -7.22$. A discrepancy of more than       
two orders of magnitude is observed ! If we use mass-loss rates higher than the one in our final model,       
a deeper blue absorption (with respect to the line center) and an intense emission    
starts to be seen, i.e., a more normal P-Cygni profile is developed, contrary to what is     
observed. On the other hand, lower $\dot{M}$'s  slowly approach the photospheric prediction by     
TLUSTY, as expected.      
      
In Figure \ref{hd216b} we present our fit to the optical spectrum of HD 216898.    
The five different spectral regions shown comprise the diagnostic lines used to derive the photospheric       
parameters (see Section \ref{method}). The effective temperature could be       
very well constrained. The \hei$/$\heii\, and \heic$/$\heiic\, ratios are well       
reproduced with a $T_{eff} = (34 \pm 1)kK$. H$\gamma$ has a reasonable fit but the       
model is somewhat stronger than the observed line. The same happens to   
H$\alpha$. Nevertheless, their wings are well matched with a log $g$ of 4.0   
$\pm$ 0.1. Details regarding this last transition are discussed later in   
Section \ref{halphasec}. Some of the most intense lines in Figure \ref{hd216b} are: 
He~{\sc i} $\lambda 4026$, H$\delta$, H$\gamma$, He~{\sc i} $\lambda 4388$, 
\hei, \heii,  C~{\sc iii} $\lambda 4647$, and C~{\sc iii} $\lambda 4650$ 
(with small contribution of O~{\sc ii}), \heiic, He~{\sc i} $\lambda 4713$, and H$\alpha$.
These same transitions are easily identified in the other objects of our
sample - and in other typical O8-9V stars - that have $vsini$ 
less than about 100 km s$^{-1}$. For rapid rotators such as $\zeta$ Oph       
and HD 216532, several profiles are broadened/blended and in some cases we    
cannot distinguish  individual transitions (see below). We highlight that the    
H$\alpha$ observed in HD 216898 is symmetric. Indeed, all model predictions    
are symmetrical for this line. However, as it is shown later, some objects     
present an asymmetric H$\alpha$ profile whose origin is not known.    
     
\subsection{HD 326329}      
        
HD 326329 is located in the NGC 6231 cluster, at the nucleus      
of the Sco OB1 association. It belongs to a resolved triple system 
(O9V+O9V+B0V; GOS Catalog), but it is probably a single star itself 
since it does not show photometric or spectroscopic variations 
(Garcia \& Mermilliod 2001). So far, no atmosphere models were 
used to analyze its spectrum.
  
In Figure \ref{hd326a} we present our final model to the far-UV and UV      
observed spectra. From the IUE continua, we have derived a E(B-V) of 0.44      
and a distance of 1.99 $kpc$, in agreement with the study of the NGC 6231      
cluster made by Baume et al. (1999). As in the case of HD 216898, a good      
agreement is found in the FUSE spectral region. The \pv,  Si~{\sc iv} $\lambda\lambda    
1122,28$, and \ciii\, lines are well reproduced. However, the IUE spectrum      
could not be entirely matched: the observed \civ\, is considerably      
deeper than in the model. We have tried several different tests to fix       
this discrepancy (e.g. changes in the $\beta$ velocity law; lower terminal   
velocities; increased turbulence), but they all turned out to be unsuccessful.      
We note that in the study of Martins et al. (2005) no such problem was      
found. The O9 dwarfs analyzed in their study present a \civ\, profile  
that was well reproduced by the models. We also emphasize that although there is only one IUE high   
resolution spectrum for HD 326329 (shown in Figure \ref{hd326a}),   
model comparisons to the other few lower resolution IUE data available   
confirm the too deep/broad feature. Interestingly, one more star of   
our sample also present this characteristic, namely, HD 216532 (see below).   
    
Despite the problem described above, we can still be confident about the    
mass-loss rate chose for HD 326329, log $\dot{M}$ = -9.22, within $\pm$0.7 dex. First, there   
is a lack of emission in the observed \civ\, P-Cygni profile, suggesting   
that $\dot{M}$ must be indeed low. 
When values for $\dot{M}$ higher than the   
one in our final model are used, a synthetic line with an intense    
(unobserved) emission and discrepant blue absorption appears   
(see Figure \ref{hd326a}; dotted line). On the other hand,  
when lower mass-loss rates are used, \civ\, slowly turn into  
photospheric features, as the ones predicted in TLUSTY.     
    
Our fit to the optical spectrum of HD 326329 is presented in      
Figure \ref{hd326b}. The He~{\sc i} and He~{\sc ii} lines     
are well reproduced with a $T_{eff} = (31 \pm 1)kK$.     
The fit to the H$\gamma$ wings gives a surface     
gravity (log $g$) of $3.9 \pm 0.1$. Both $T_{eff}$ and     
log $g$ were previously estimated by Mathys et al. (2002)      
who found 31700$K$ and 4.6, respectively, by using $uvby\beta$     
photometry. Although our derived temperature is compatible with their value,     
their higher log $g$ is not supported by our analysis. In fact,     
a log $g$ much higher than 4.0 is not expected for O stars     
(see e.g. Vacca et al. 1996; Martins et al. 2002).     
Although a fairly good agreement is found in several parts of    
the optical spectrum, we could not achieve a good fit to H$\alpha$.    
This line is weaker than in the model and is also asymmetric.    
Its blue wing has a ``bump'' which distorts the profile.   
This is not seen in the other hydrogen lines.    
We have not used H$\alpha$ to derive any stellar or wind parameters    
of the objects of our sample. However, we later discuss the effects    
that different mass-loss rates have on this line (see Section \ref{halphasec}).     
             
\subsection{HD 66788}        
       
Figure \ref{hd66a} shows our fits to the ultraviolet spectrum of HD 66788. 
The \civ\, feature is reproduced with a mass-loss rate of log $\dot{M}$ =
-8.92. This value is much lower than the one predicted, 
log $\dot{M}_{Vink} = -6.95$. The reddening and distance derived are     
E(B-V) = 0.22 and d = 4.8kpc. Although large, this distance is consistent     
with estimates encountered in the literature (Reed 1993;     
Kaltcheva \& Hilditch 2000).     
     
In Figure \ref{hd66b} we show our fit to the optical spectrum. 
The features found in HD 216898 and HD 66788 are very similar, 
suggesting similar photospheric parameters (see Section \ref{hd216sec}). 
Indeed, from our fits to the H$\gamma$ we derive a log $g = 4.0 \pm 0.1$, 
and from the \hei\, and \heii\, lines we obtain a $T_{eff} = (34 \pm 1)kK$, 
as in HD 216898. Their rotational velocities are also practically the same, 
about 60 km s$^{-1}$. We note that the H$\alpha$ line in HD66788 is 
slightly asymmetric. Although the overall agreement is reasonable, 
its blue wing has a weak absorption not predicted by the model.
     
\subsection{$\zeta$ Oph}       
       
$\zeta$ Oph (also known as HD 149757) is a well studied runaway    
star which presents several interesting characteristics. It has been known for a        
long time to present different kinds of spectral        
variability, such as discrete absorption components        
in the UV (DACs), emission line episodes, and line        
profile variations (LPVs) (see e.g.         
Howarth et al. 1984; Reid et al. 1993; Howarth et al. 1993;        
Jankov et al. 2000; Walker et al. 2005). This object        
presents also a very high rotational velocity: $vsini$        
values around 400 or even 500 km s$^{-1}$ were already        
reported in the literature (Walker et al. 1979;        
Repolust et al. 2004).  
       
A reason to include $\zeta$ Oph in our sample,        
is that it is considered to have a $\dot{M}$ estimate        
based on H$\alpha$ which is in good agreement         
with the predictions of the radiative wind theory        
(Mokiem et al. 2005; 2007). Thus, it is interesting     
to check if a fit from the UV to the optical     
(including H$\alpha$) can be achieved with CMFGEN     
using a low $\dot{M}$. We stress however, that the     
very high rotational velocity presented by     
$\zeta$ Oph can bring some problems to the analysis.        
First, fast rotation can distort the shape and     
induce temperature and gravity variations     
through the stellar surface (see e.g. Fr\'emat et al. 2005).     
Furthermore, it might imply that a stellar wind is not     
spherically symmetric. In such cases, 1D unified      
atmosphere models should be considered as an approximation.     
Another difficulty is that several features in the     
spectrum are broadened and sometimes blended. The analysis      
of diagnostic lines thus can present a greater difficulty     
than in the case of slower rotators.        
  
\begin{table*}        
\caption{Stellar and Wind Parameters.}        
\begin{tabular}{l|ccccc}        
\hline\hline        
Star                  & HD 216898       & HD326329      &  HD 66788       & $\zeta$ Oph     & HD 216532 \\  \hline        
Spec. type             & O9IV, O8.5V    & O9V           &  O8-9V          & O9.5Vnn         &  O8.5V((n))    \\        
log g                  &  4.0 $\pm$ 0.1 & 3.9 $\pm$ 0.1 &  4.0 $\pm$ 0.1  & 3.6 $\pm$ 0.2   &  3.7 $\pm$ 0.2  \\        
T$_{eff}$ (kK)          &  34 $\pm$ 1    & 31 $\pm$ 1    &  34 $\pm$ 1     & 32  $\pm$ 2     & 33 $\pm$ 2    \\        
vsini (km s$^{-1}$)     &  60            & 80            &  55             & 400             & 190           \\         
log L/L$_\odot$         & 4.73 $\pm$ 0.25 & 4.74 $\pm$ 0.10 &  4.96 $\pm$ 0.25 & 4.86 $\pm$ 0.10 &  4.79 $\pm$ 0.25  \\        
R$_\star$ (R$_\odot$)   &$6.7^{+2.3}_{-1.7}$& $8.0^{+1.1}_{-1.0}$&$8.7^{+3.0}_{-2.2}$ &$9.2^{+1.7}_{-1.4}$&$7.5^{+2.7}_{-2.0}$   \\    
M$_\star$ ($M_\odot$)   &  17$^{+15}_{-8}$  &   19$^{+8}_{-6}$ & 26$^{+23}_{-12}$  & 13$^{+10}_{-6}$  &  12$^{+14}_{-6}$    \\  
v$_\infty$ (km s$^{-1}$) &  1700 $\pm$ 500   & 1700 $\pm$ 500  &  2200 $\pm$ 500&  1500 $\pm$ 500   &  1500 $\pm$ 500        \\        
log $\dot{M}$          &  -9.35 $\pm$ 0.7   & -9.22 $\pm$ 0.7 &  -8.92 $\pm$ 0.7 & -8.80 $\pm$ 0.7  &  -9.22 $\pm$ 0.7         \\        
log $\dot{M}_{Vink}$    &  -7.22         & -7.38         &  -6.95           & -6.89           &  -6.92         \\        
\hline\hline        
log $L_x/L_{Bol}$      &  -7.00          & -6.69          &  -7.06          & -7.31           & -7.00            \\        
log L/$c^2$            &  -8.42         & -8.41         &  -8.19           & -8.29           &  -8.36         \\        
E(B-V)                &  0.7            & 0.44           &   0.22          & 0.36            &  0.72             \\        
R                     &  3.5            & 3.1            &   2.8           & 2.9             &  3.5            \\       
distance (pc)         &  1100           & 1990           &   4800          & 146             &  1150          \\       
\hline        
\end{tabular}        
\label{results}      
\end{table*}

Our final model and the IUE spectrum of $\zeta$ Oph are shown in      
Figure \ref{zeta1}. A distance of 146pc and an E(B-V)=0.36     
was used for the continuum fit. Due to the high $vsini$     
involved, the photospheric iron forest is considerably broadened.    
Despite weak, \nv\, can be easily distinguished in     
the spectrum, as well as the \ciii\, and the \siiv\, lines.     
As it is clear, the model is successful     
to reproduce the main UV features observed, but some discrepancies     
can be noted. The \civ\, line for example, could not be     
matched in detail; the synthetic line is slightly more intense     
and narrower than the observed one. A model with a      
mass-loss rate lower than in our final model does not solve    
the problem and has also a drawback in the \nv\, fit.    
Also, a higher mass-loss rate implies in a stronger     
emission and blue absorption of the P-Cygni     
profile, in contrast with the observations (see Figure \ref{zeta1}; dotted   
line). The rate used, log $\dot{M} = -8.80$, should be certain within a    
factor of three (i.e. $\sim$0.5 dex). This value is much lower than the one    
derived in the study of Mokiem et al. (2005). Using the FASTWIND code, and   
based on H$\alpha$, these authors derive log $\dot{M}(H\alpha) = -6.85$.    
Our value is also much lower than the one predicted according to Vink    
et al. (2000), log $\dot{M_{Vink}} = -6.89$. We come back to this question later in Section \ref{mdotlateo}.    
Some other smaller discrepancies can be also seen in the \siiv\,    
line fit and on the $\sim 1550-1650$\AA\, interval. Despite our efforts, they could     
not be sorted out. Due to the very high rotational velocity presented by $\zeta$     
Oph, future studies using 2D atmosphere models might be very useful to address some of     
these problems.   

In Figure \ref{zeta2} we present our model to the optical spectrum. 
As we mentioned earlier, a high rotational velocity     
convolution is necessary to fit the broad features observed     
($vsini \sim$ 400 km s$^{-1}$). Despite the high rotation,     
the effective temperature and surface gravity could be derived,     
but with a higher uncertainty than in the other objects of our sample.     
From the  He~{\sc i} and He~{\sc ii} lines, we have obtained a     
$T_{eff} = (32 \pm 2)kK$. From the H$\gamma$ profile we estimate a     
low surface gravity compared to the other objects: log $g = 3.6 \pm 0.2$.     
It should be kept in mind that this value should be regarded as     
an {\it effective gravity}, i.e., the gravity attenuated by     
rotation. A ``centrifugal correction'' can be applied in order     
to obtain the {\it true} gravity, according to the equation     
log $g_{true} = $ log $(g_{eff} + vsini^2/R_{\star})$ (see     
details in Repolust et al. 2004). For $\zeta$ Oph, we find     
a log $g_{true}$ of $\sim 3.8$. This correction however,     
neglects any distortion in the stellar shape. A treatment     
allowing the oblateness of the star was presented by     
Howarth \& Smith (2001) for three fast rotators, including     
$\zeta$ Oph. These authors derived a polar (equatorial)     
gravity of $\sim$4.0 ($\sim$3.6), a polar (equatorial) radius of 7.5R$_\odot$    
(9.1R$_\odot$), and an effective temperature of $\sim 34kK$.     
The radius obtained from our $T_{eff}$ and $L_\star$ is 9.2R$_\odot$.     
Therefore, our stellar parameters for $\zeta$ Oph are compatible     
to the equatorial values found by Howarth \& Smith (2001). They     
also show agreement with other works in the literature     
(Repolust et al. 2004; Mokiem et al. 2005; Villamariz \& Herrero 2005).       

Turning our attention to H$\alpha$, we note that this line and     
a few others were observed to present emission episodes (in the     
form of double peaks) on different occasions (see e.g. Niemela     
\& Mendez 1974; Ebbets 1981). However, in general, H$\alpha$ is     
observed to be a symmetric absorption, with an equivalent width     
($W_\lambda$) of $(2.7 \pm 0.4)$\AA, and with a central depth of     
about 0.8 of the continuum intensity. These characteristics can be     
considered to represent $\zeta$ Oph's {\it quiescent} spectrum     
(see Reid et al. 1993 and references therein). In our case, from the     
observed line we measure a $W_\lambda$ of $(3.0 \pm 0.3)$\AA\, and the     
deepest part of the absorption is also at $\sim$0.8 of the continuum level,     
confirming a {\it quiescent}-like, i.e., normal spectrum. However, the     
profile presents narrower wings if compared to the one in our    
final model (see Figure \ref{zeta2}, bottom). This problem can be    
also perceived in the model presented by Mokiem et al. (2005) using    
the FASTWIND code (see their Fig. 12). A decrease in the $vsini$    
to 350 km s$^{-1}$ improves the fit to the wings, but the line    
center gets slightly deeper than observed. Such lower $vsini$ has    
also a negative effect in other optical lines such as H$\delta$ and H$\gamma$.       

\subsection{HD 216532}             

HD 216532 is a relatively fast rotator which belongs to the 
Cepheus OB3 association. According to Howarth et al. (1997)     
its $vsini$ is $\sim$190 km s$^{-1}$, a value that we confirm     
from our model fits. Leitherer (1988) have estimated its     
stellar parameters and derived an upper limit for the      
mass-loss rate from H$\alpha$, log $\dot{M} < -7.08$.
The present study is the first to analyze quantitatively     
its far-UV to optical spectra.
    
Our fit to the far-UV and UV data is presented in Figure \ref{hd216532a}. 
For the E(B-V) and the distance, we inferred 0.72 and 1.15kpc, respectively. 
In the FUSE region, although the overall agreement is fair, some discrepancies can be noted.     
The observed \pv\, and Si~{\sc iv} $\lambda \lambda 1122,28$ lines are deeper     
than in the model. Although we can see that the synthetic      
continuum is somewhat higher than the observed, the     
problem remains if we normalize the spectrum locally.     
Regarding the IUE spectrum, as in HD 326329, we could     
not fit the \civ\, line. Compared to the profiles found     
in HD 216898 and HD 66788, the \civ\, line in HD 216532     
is broader and deeper. Nevertheless, the mass-loss rate of    
our final model, log $\dot{M} = -9.22$, should be certain    
within about $\pm$0.4 dex. Higher values result in P-Cygni    
profiles whose emissions are too strong, and this is not    
observed. An example is shown in Figure \ref{hd216532a}    
(see the dotted line). As we already mentioned, lower mass-loss rates slowly    
change the \civ\, feature into two photospheric absorptions.    

In Figure \ref{hd216532b} we present the observed optical     
spectrum of HD 216532 along with our final model. It     
can be readily noted that the lines are broader if compared     
to the ones in HD 216898, HD 326329, and HD 66788,     
reflecting the relatively high $vsini$ of this object, $\sim$190 km s$^{-1}$.     
With an effective temperature of $\sim$33kK, we achieve a good    
match to the observed \hei\, and \heii\, lines. To fit the H$\gamma$    
profile, we have used a log $g$ of $3.7$.    

\begin{figure*}      
\center    
\includegraphics[width=16cm,height=8cm]{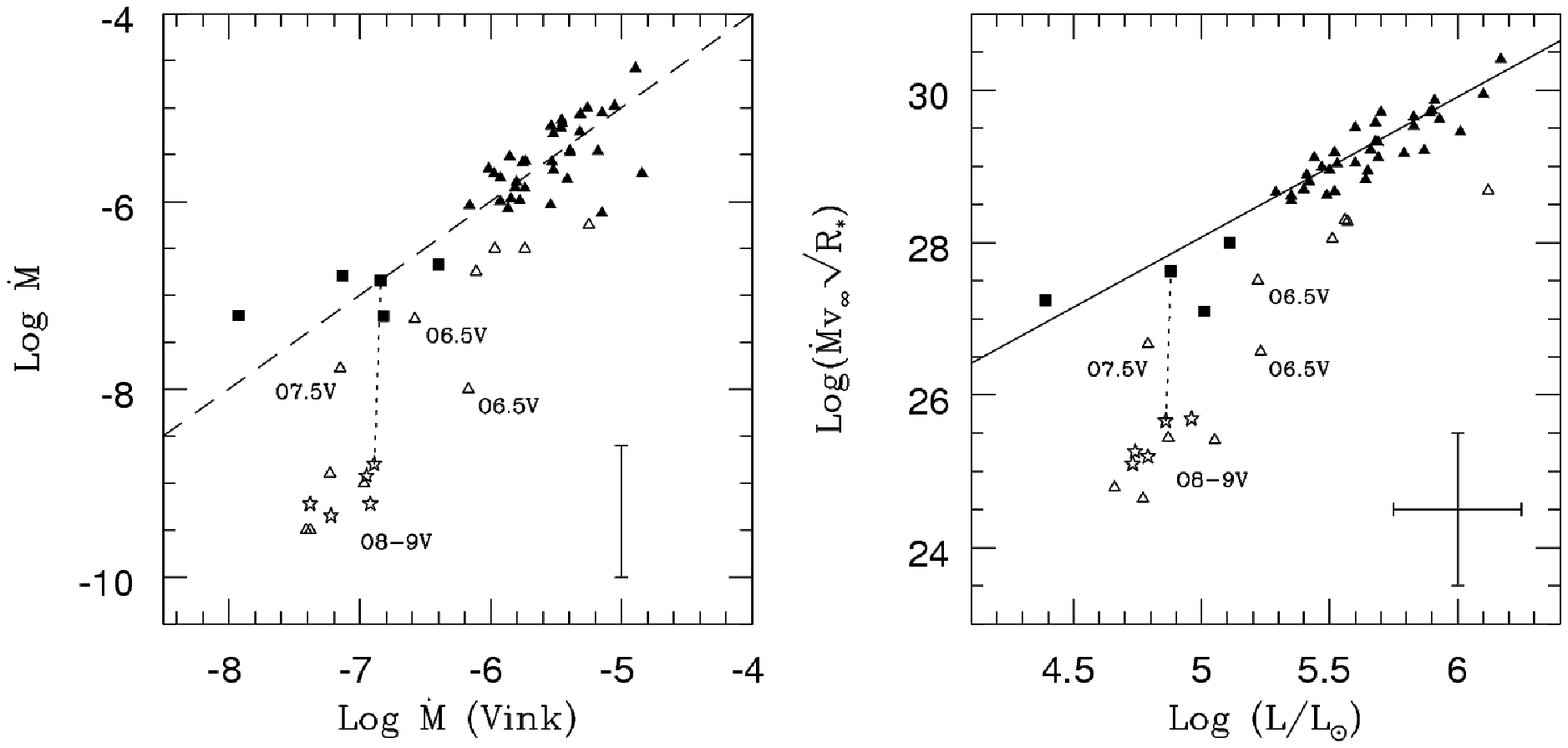}        
\caption{Comparison of our results to previous ones in the literature and    
theoretical predictions. The objects of our sample are represented by star symbols;    
the O dwarfs studied by Martins et al. (2005) by open triangles;     
the O and B stars compiled by Mokiem et al. (2007) that have    
log $L_\star/L_\odot \lesssim 5.2$ are represented filled squares and    
the ones with higher luminosities by filled triangles.   
Left panel: mass-loss rates compared to theoretical predictions    
($\dot{M}_{Vink}$). The dashed line indicates a one-to-one relation.   
Right panel: modified wind momentum-luminosity relation (WLR). The solid    
line represents the best fit to the data in Mokiem et al. (2007).   
The dotted lines link the results obtained in this paper and in Mokiem  
et al. (2005) regarding the star $\zeta$ Oph.   
Low luminosity dwarfs (log $L_\star/L_\odot \lesssim 5.2$) in    
this paper and in Martins et al. (2005) are explicitly indicated by their    
spectral types (see text for more details).}        
\label{results_plot}        
\end{figure*}         

\section{Analysis of the Results:}      
\label{summary} 

In this section we summarize the results of our spectral analysis and     
make a comparison to previous works and theoretical predictions.    
The stellar and wind parameters obtained for each star of our     
sample are presented in Table \ref{results}. For comparison, we also     
list the theoretical mass-loss rates computed following the recipe of Vink     
et al. (2000), $\dot{M} _{Vink}$. Additional parameters such as the     
reddening and the distances are shown in the lower part of the table.    
Regarding the X-rays luminosities, we recall that for HD 216898, HD 66788,   
and HD 216532, we have fixed log $L_X/L_{Bol}$ at the canonical value,   
i.e. at $\sim -7.0$. For HD 326329 and $\zeta$ Oph, we chose   
$L_X/L_{Bol}$ ratios close to the ones recently observed   
(Oskinova 2005; Oskinova et al. 2006; Sana et al. 2006).     

Overall, the stellar properties of our sample are quite homogeneous.    
The effective temperatures obtained range from $\sim$30-34$kK$, and     
the radii are between $\sim$7-9$R_\odot$. Regarding the surface     
gravities, the most different (lowest) values are presented by     
HD 216532 and $\zeta$ Oph. In both stars, the log $g$      
measured should be interpreted as effective, i.e., they are     
attenuated by their fast rotation. These physical properties    
show a fair agreement with the latest calibration of     
Galactic O star parameters, regarding the O8-9V spectral types     
(see Martins et al. 2005b).    
    
In order to better analyze the stellar and wind parameters     
derived, we also present our results in Figure \ref{results_plot},   
in two different ways: the mass-loss rates are compared to the theoretical    
predictions of Vink et al. (2000); and the wind parameters    
are used to construct the modified wind momentum luminosity    
relation (WLR): $(\dot{M}v_\infty \sqrt{R_\star}) \times L_\star$.  
In each of these plots, we include the results of Martins  
et al. (2005) regarding eleven O dwarfs, as well as the data 
gathered by Mokiem et al. (2007) regarding O and B dwarfs, giants, and
supergiants (see their Table A1). In order to simplify the analysis and    
exclude metallicity effects, only Galactic objects are considered.
    
In the plots presented in Figure \ref{results_plot}, we can     
perceive the same basic result: the stars of our sample gather    
around the same places occupied by the four O8-9V stars studied    
in Martins et al. (2005), namely, $\mu$ Col, AE Aur, HD 46202, and HD 93028.    
This means that: (i) we have also found a discrepancy of roughly    
two orders of magnitude between the measured and the predicted    
mass-loss rates for our programme stars; (ii) the modified  
wind momentum-luminosity relation indeed shows a breakdown or  
a steepening below log $L_\star/L_\odot \sim 5.2$.  
   
Regarding the brighter objects, i.e. with log $L_\star/L_\odot > 5.2$,    
they follow reasonably well the theoretical expectations regarding    
the mass-loss rate. We remind that the four most luminous    
early-type dwarfs analyzed in the work of Martins et al. (2005) have clumped    
mass-loss rates. Thus, although they fall somewhat below    
the theoretical line in the log $\dot{M}$ x log $\dot{M}_{Vink}$    
plot and below the fit to the data of Mokiem et al. (2007), they    
do not present a significant discrepancy if clumping is    
neglected\footnote{We recall that the theoretical    
models of Vink et al. (2000) do not include clumping.}. 
 
The results derived by us and Mokiem et al. (2005) regarding the star    
$\zeta$ Oph are connected in Figure \ref{results_plot}. The large difference  
observed is due to the different mass-loss rates derived. While we have   
log $\dot{M} = -8.80$ from the \civ\, line, Mokiem et al. (2005) derived log $\dot{M} = -6.84$    
based on H$\alpha$, using the FASTWIND code. The other stellar and wind parameters    
determined in their study and ours are very similar. The radius, $T_{eff}$, and    
log $L_\star/L_\odot$ from this (their) paper are 9.2 (8.9) $R_\odot$,    
32.1 (32) kK, and 4.86 (4.88), respectively. Regarding the wind terminal  
velocity, we (they) obtain 1500 (1550) km s$^{-1}$.
   
As we mentioned above, the objects of our sample and the four O8-9V stars    
analyzed in Martins et al. (2005) occupy about the same locus in the panels    
shown in Figure \ref{results_plot}. It is interesting however to    
analyze their results for other stars having also low luminosities, i.e.,    
with log $L_\star/L_\odot \lesssim 5.2$. Two objects in    
Martins et al. (2005) have a log $L_\star/L_\odot$ of about 5.2    
($\pm 0.2$ dex) and are classified as O6.5V stars: HD 93146 and HD 42088.    
Another object included in their study is HD 152590, an O7.5V with    
a log $L_\star/L_\odot \sim 4.8$. They fall relatively far from the O8-9V group    
in the plots in Figure \ref{results_plot}. Contrary to the other four earlier O  
dwarfs, clumping was not used in their analysis. These three examples are  
suggestive that in stars of the spectral types O6.5V, O7V, and O7.5V, a  
decrease of the wind strength starts to be seen, culminating in the O8-9V classes.    
       
Also at low luminosities are the following objects analyzed by Mokiem et    
al. (2005): Cyg OB2\#2 (B1 I), $\tau$ Sco (B0.2V), 10 Lac (O9V),    
and HD 217086 (O7Vn) (see Figure \ref{results_plot}; filled squares).    
They deserve special attention. In their case, their mass-loss rates clearly    
present better agreement to the theoretical predictions and also allow them    
to follow relatively well the WLR. Interestingly,     
their $\dot{M}$'s were obtained from the H$\alpha$ line.    
Given these and our UV based results, the origin of the weak wind problem    
can be thought to reside in the different mass-loss rates diagnostics   
employed. In fact, Mokiem et al. (2007) have suggested that the problem could    
be due to uncertainties in the UV method. We address and discuss    
these questions during the rest of the paper.    
   
We emphasize that there are several other O8-9V stars with UV spectra
 similar to the ones found in our sample. Thus, it is likely that they 
possess similar wind properties, i.e., that the same basic results we have 
obtained will be achieved for them if we use the same methods of analysis.  
 
\section{The Carbon Abundance}       
\label{carbona}

\begin{figure*}      
\center   
\includegraphics[width=18cm,height=20cm]{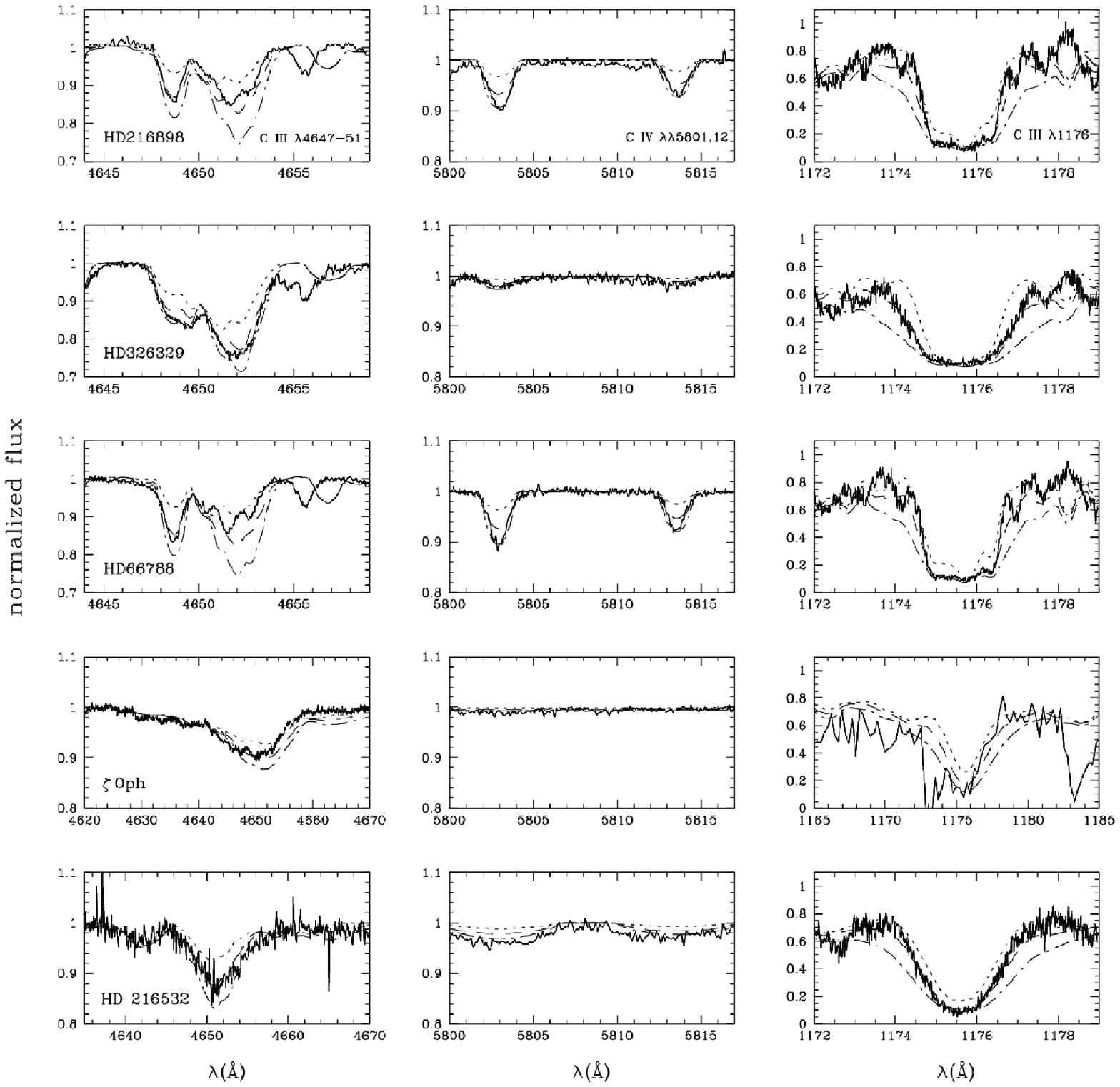}        
\caption{Determination of the carbon abundance in our programme stars.   
The observed spectra are indicated by solid line lines. The models shown have:    
20\% solar (0.2$\epsilon_{C_{\odot}}$; dotted line), solar ($\epsilon_{C_{\odot}}$;   
dashed line), and 3x the solar carbon abundance (3$\epsilon_{C_{\odot}}$;dashed-dotted line).}        
\label{carb}        
\end{figure*}     
       
From the models, it can be easily seen that changes in the carbon 
abundance ($\epsilon_{C}$) affect the \civ\, profile. In fact, by choosing  
different combinations of $\dot{M}$ and $\epsilon_{C}$ we can  
get satisfactory fits to this feature, as long as  
its optical depth is kept constant, i.e., at the {\it observed} value.  
This can be seen more directly in terms of the Sobolev optical depth,  
which is proportional to the mass-loss rate, the C~{\sc iv} ionization  
fraction ($q_{CIV}$), and the abundance: $\tau_{CIV} \propto \dot{M}\,q_{CIV}\,\epsilon_C$ 
(see for example Lamers et al. 1999). In the context of the weak wind 
problem, this suggests that we could fit \civ\, using models with  
mass-loss rates close to the values predicted by the radiative wind  
theory (Vink et al. 2000) and low amounts of carbon. We conclude  
below however, that this cannot be the case.    
   
It is clear that an accurate carbon abundance determination in     
late O dwarfs is needed to disentangle $\dot{M}$ and $\epsilon_C$.    
In order to achieve this goal, we have built a small grid of models with the following     
$\epsilon_C/\epsilon_{C_\odot}$ ratios: 0.2, 0.5, 1.0, 1.5, and 3 (mass fractions).     
Thereafter, we have analyzed the behavior of the following     
photospheric transitions: C~{\sc iii} $\lambda$4647-4651, $\lambda$1176,      
and C~{\sc iv} $\lambda\lambda$5801,12. Although these lines do  
respond to changes in $\epsilon _C$, an accurate determination of  
the carbon abundance is not feasible at the moment.

In Figure \ref{carb} we present the fits obtained   
using models with $\epsilon_C = 0.2\epsilon_{C_\odot}$     
(20\% solar), $\epsilon_C = \epsilon_{C_\odot}$ (solar), and $\epsilon_C = 3\epsilon_{C_\odot}$     
(3$\times$ solar). Although our grid is more complete, we illustrate only   
extreme $\epsilon_C$ values (besides the solar one) to emphasize the  
discrepancies. From Figure \ref{carb}, we can note that: (i) when an abundance     
of 0.2$\epsilon_{C_\odot}$ is used, we cannot fit any of the lines     
shown; (ii) when a 3$\epsilon_{C_\odot}$ value is used we cannot     
fit properly the \ciii\, line, because a large/broad wing     
appears in the synthetic line and this is not observed;     
(iii) the 3$\epsilon_{C_\odot}$ abundance is also not favored from     
the fits to C~{\sc iii} $\lambda$4647-4651 in HD 216898 and     
HD 66788, and hints to be an excess also in HD326329, $\zeta$ Oph,     
and HD 216532.  From (i) we can conclude that our sample must have     
a $\epsilon_C > 0.2\epsilon_{C_\odot}$. On the other hand, from (ii)     
and (iii) we are inclined to conclude that $\epsilon_C < 3\epsilon_{C_\odot}$.    
Regarding the C~{\sc iv} $\lambda\lambda$5801,12 lines, we see that     
a $3\epsilon_{C_\odot}$ upper limit is also suggested from the fits.     
      
From our analysis, we estimate the following abundance     
range for the O8-9V stars in our sample: $0.5 \lesssim \epsilon _C / \epsilon _{C_{\odot}} \lesssim 2$.     
Although the uncertainty is large, it is not large enough to     
allow the use of mass-loss rates that are consistent with the     
values predicted by theory ($\dot{M}_{Vink}$). A simple test     
show that if an abundance as low as 0.2$\epsilon_{C_{\odot}}$ (not     
favored by our models) is used along with $\dot{M}_{Vink}$,     
the model still present a very intense \civ\, P-Cygni profile     
which is not seen in any O8-9V stars. Hence, although the     
uncertainty in the carbon abundance adds to the uncertainty     
in $\dot{M}$, it cannot cause the weak wind problem.     
The abundance range obtained does not leave enough room for 
the mass-loss rates to be consistent with $\dot{M}_{Vink}$. 
A similar conclusion was previously achieved by Martins et al.     
(2005), who quantified the error on the mass-loss rate     
due to the uncertainties in the CNO abundances to be     
about 0.3dex.           

\section{Mass-loss Rates in Late-Type O Dwarfs:}        
\label{mdotlateo}   
      
\begin{figure*}        
\center   
\includegraphics[width=16cm,height=9cm]{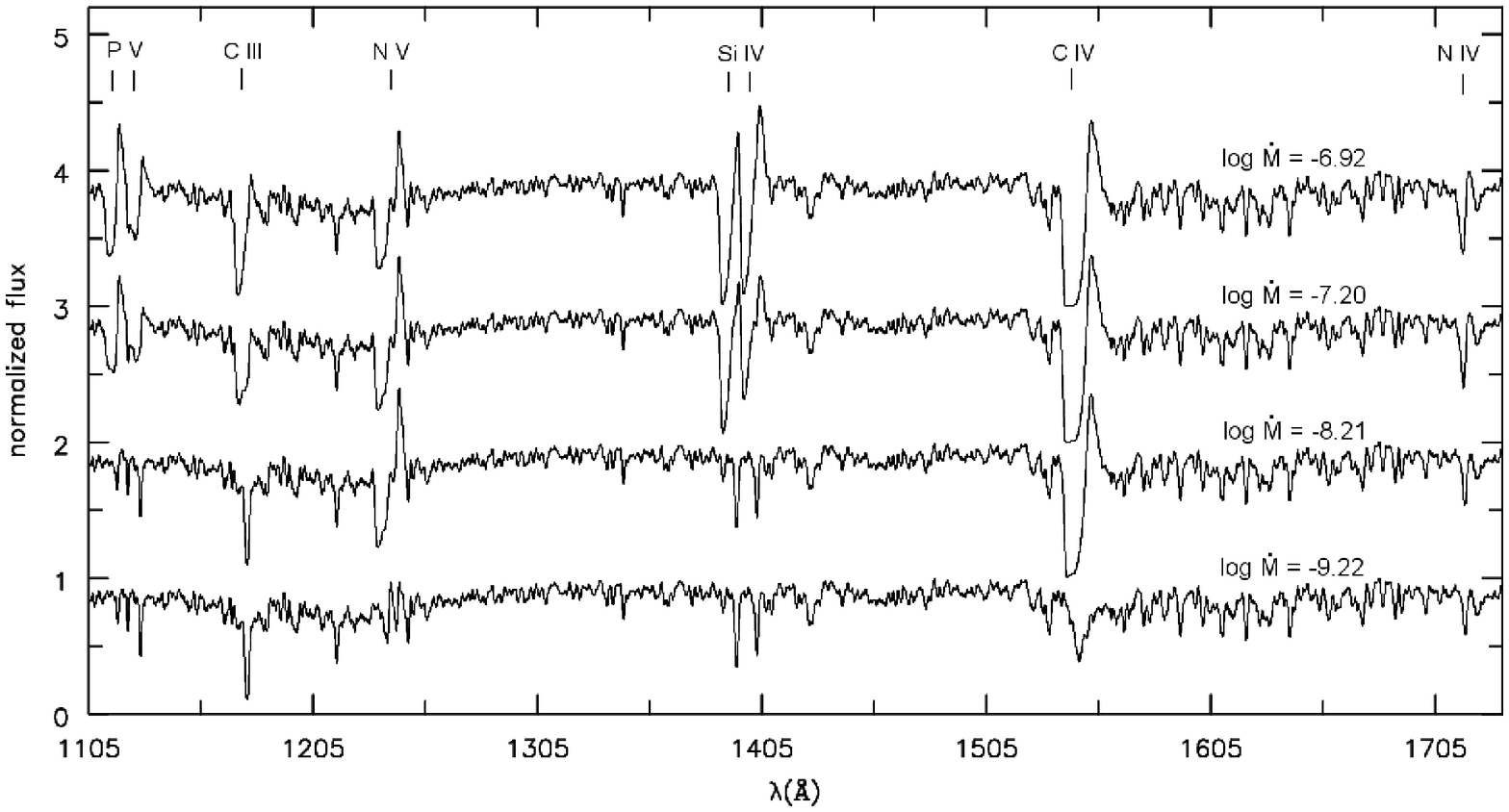}        
\caption{$\dot{M}$-sensitive lines in the far-UV and UV regions.     
A $vsini$ convolution of 200 km s$^{-1}$ was applied in each spectrum.  
Other physical parameters (e.g. $T_{eff}$ and $v_\infty$) are held    
fixed at the values derived for HD 216532 (see Table \ref{results}).   
The models were normalized and vertically displaced for clarity.}        
\label{sensitive_lines}        
\end{figure*}            
     
\begin{table*}        
\caption{Mass-loss Rates and Upper Limits Derived from Different Transitions (log units).}        
\begin{tabular}{lccccccc}       
\hline\hline        
Star               & $\dot{M}$ (C IV) & $\dot{M}$ (C III) & $\dot{M}$ (N V) & $\dot{M}$ (N IV) & $\dot{M}$ (P V) & $\dot{M}$ (Si IV) & $\dot{M}$ (Vink)   \\  \hline        
HD 216898          & -9.35 & $<$-7.47 & $<$-8.46 & $<$-7.80 & $<$-7.96 & $<$-7.59 & -7.22 \\        
HD 326329          & -9.22 & $<$-7.73 & $<$-8.22 & $<$-7.49 & $<$-7.73 & $<$-7.96 & -7.38 \\         
HD 66788           & -8.92 & $<$-7.22 & $<$-8.38 & $<$-7.52 & $<$-7.52 & $<$-7.22 & -6.95 \\        
$\zeta$ Oph        & -8.80 & $<$-7.40 & $<$-8.15 & $<$-6.89 & $<$-7.40 & $<$-7.70 & -6.89 \\      
HD 216532          & -9.22 & $<$-7.45 & $<$-8.21 & $<$-7.57 & $<$-7.45 & $<$-7.70 & -6.92\\            
\hline        
\label{uppermd}       
\end{tabular}        
\end{table*}   
 
\subsection{Other $\dot{M}$ Diagnostics:}       
\label{othermdot}      
    
In most cases, the \civ\, resonance doublet is the only diagnostic available to 
establish mass-loss rates in late-type O8-9V stars, which brings uncertainties to the 
discussion of the weak wind problem. In order to overcome this 
situation, we have explored an alternative approach to estimate 
$\dot{M}$ values.
 
We have proceeded in the following manner: for each object of     
our sample, we have started with the $\dot{M}$ determined from the best     
fit to \civ\, (the final model). Then, we have gradually increased this     
parameter until reach the value predicted by the radiative wind theory     
- $\dot{M}_{Vink}$ - which for each star is a     
function of $T_{eff}$, mass and luminosity. We observed that     
as soon as we have used values higher than in our final    
models (yet considerably lower than $\dot{M}_{Vink}$), different     
wind profiles started to appear which are not observed.    
We have thus used this finding to establish upper limits        
on $\dot{M}$ from different transitions, as described below. 
 
First, in Figure \ref{sensitive_lines}, we illustrate the $\dot{M}$  
diagnostics found by following the procedure aforementioned.  
Besides \civ, the lines that show significant changes are:     
\pv, \ciii, \nv, \siiv, and \niv. When $\dot{M}$ is low     
(i.e. $\sim 10^{-10}$-$10^{-9}$ M$_\odot$ yr $^{-1}$), \nv\,    
is generally weak (or absent) and the other lines  
are essentially photospheric. However, when $\dot{M}$ is  
increased they gradually have their profiles modified and  
some develop a very intense P-Cygni profile when $\dot{M}_{Vink}$  
is used (see for instance \siiv). 
 
The upper limits ($\dot{M}_{upper}$) for these $\dot{M}$-sensitive lines were  
obtained in different ways. For \pv, \ciii, and \siiv, we have observed when  
the profiles started to get filled by wind emission and/or a blue  
absorption started to appear, deviating from the (purely photospheric) 
observed ones. For \nv, we used the strength of the whole P-Cygni profile, which  
generally is weak or absent in the observations. The \niv\, line has a different  
response. Instead of getting filled in its center, an asymmetric blueward  
absorption profile is formed. We have then determined $\dot{M}_{upper}$  
when a significant (blue) displacement from the observed feature was reached.  
Such peculiar behavior indicates that this line is formed  
in the very inner parts of the stellar wind. In the study done by Hillier et  
al. (2003), a similar situation is found in the homogeneous model  
for the O7 Iaf+ star AV 83, specially regarding S~{\sc v} $\lambda$1502  
(see their Fig. 10). 

The results obtained for each object of our sample are summarized  
in Table \ref{uppermd}. The \civ\, mass-loss rates indicated are the ones  
used in our best fits presented in Section \ref{analysis}. The predicted  
mass-loss rates are also shown for comparison. Below we present the  
models used to construct Table \ref{uppermd} and comment on each object.  
    
\begin{figure}        
\includegraphics[width=9cm,height=9cm]{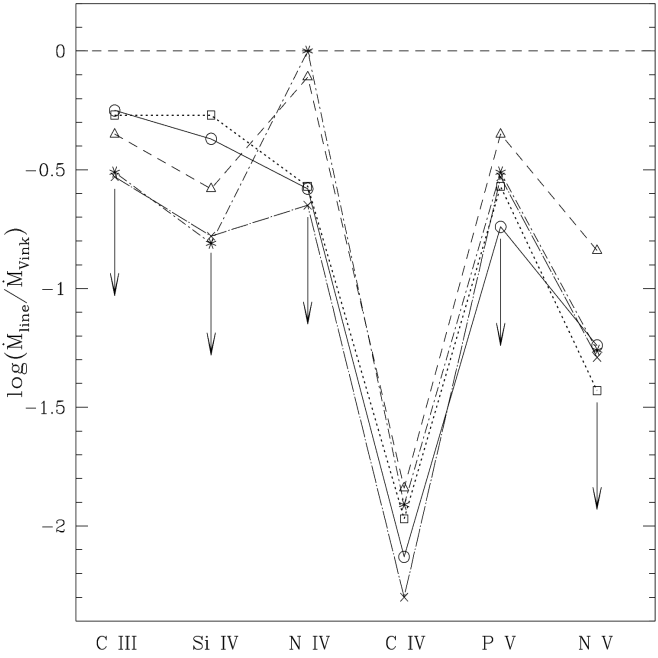}        
\caption{Mass-loss rates upper limits derived from different transitions in    
the ultraviolet. The stars are: HD 216898 (circled+solid line), HD 326329 (triangles+dashed 
line), HD 66788 (squares+dotted line), $\zeta$ Oph (asterisk+dashed-dotted 
line), and HD 216532 (crosses+short-long dashed line). Arrows indicate that 
the points represent upper limits.}        
\label{upperplot}        
\end{figure}        
    
\begin{itemize}    
    
\item {\bf HD 216898:} In Figure \ref{mdot_hd216898} we     
show the far-UV and UV observed spectra of this object     
along with our final model, the models that determine      
the upper limits in $\dot{M}$ from each line ($\dot{M}_{upper}$; not necessarily the     
same in each panel), and the model with the Vink's     
mass-loss rate. We can clearly note from Figure \ref{mdot_hd216898} that:     
(i) the final model presents the best fit to the observed     
spectra; (ii) the model with $\dot{M}_{Vink}$ presents the largest     
discrepancy; and (iii) the models with the upper     
limits for the mass-loss rate do not present satisfactory fits.     
This is also valid for the other objects discussed below.    
We note that when we have increased $\dot{M}$,     
the \nv\, line quickly turned to a P-Cygni profile     
and remained essentially unchanged even when we     
have reached $\dot{M}_{Vink}$.\\    
    
\item {\bf HD 326329:} An analogous plot as shown for HD 216898     
is presented in Figure \ref{mdot_hd326329} for this object.     
We observed that the \niv\, transition is not so sensitive     
to $\dot{M}$ in this case. Therefore, the upper limit for $\dot{M}$     
derived from this line is quite close to $\dot{M}_{Vink}$.    
For the other lines the situation was much clearer, i.e.,     
the upper limits could be easily obtained.\\    
    
\item {\bf HD 66788:} In Figure \ref{mdot_hd66788} we show     
the case of HD 66788. In general, the same line trends found     
in HD 326329 and HD 216898 are observed. We had no problems     
in the determination of $\dot{M}$ upper limits from     
each line.\\    
   
\item {\bf $\zeta$ Oph:} The observed spectra and     
models are shown in Figure \ref{mdot_zeta}. Regarding     
\niv, we see that our final model does not present an     
adequate fit to the observed line, which is considerably     
deeper. In fact, none of the explored $\dot{M}$ values could     
fit this feature. By increasing the mass-loss rate however,     
we noted that the absorption still went blue-shifted  
(as in the other objects). We conservatively assume a $\dot{M}$  
upper limit from this line equals to the Vink's mass-loss.  
Another difficult situation is presented by \nv.  
First, unexpectedly, the model using $\dot{M}_{Vink}$     
(i.e. with the strongest mass-loss rate) has a synthetic     
P-Cygni emission less intense than the one in the upper limit     
model. A slightly stronger emission is however found at     
$\sim$1249\AA. In order to choose the upper limit model     
shown in the figure, we have focused in the absorption    
part of the P-Cygni profile. The theoretical line is deeper    
than observed, specially around 1235\AA. We have not    
considered in the analysis the two (too) deep absorptions    
observed as part of the wind profile (see the two marks    
in Figure \ref{mdot_zeta} in this line). These    
features present a large variability in $\zeta$ Oph    
(mainly the bluest one), which is interpreted as being due    
to DACs (Howarth et al. 1993; see their Figure 2). The    
situation for \pv\, and \siiv\, is     
somewhat simpler. Although our final model does not present     
a very good match to the observed lines, it correctly     
predicts them in absorption. The same does not happen with      
the upper limit and Vink's mass-loss models. \\    
    
\item {\bf HD 216532:}  In Figure \ref{mdot_hd216532} is     
shown the models and observed spectra of HD 216532.     
We note that the \ciii, \pv, and \siiv\, lines     
present a well developed P-Cygni profile with $\dot{M}_{Vink}$.     
Regarding the \pv\, lines, the final model fit is not     
deep enough. Nevertheless, an upper limit for $\dot{M}$ can     
be derived without difficulty as these lines      
turn to emissions when $\dot{M}$ is increased, and this is    
not what is observed.   
    
\end{itemize}    
  
In Figure \ref{upperplot} we explore the results presented     
above in a graphical way. The mass-loss rates obtained    
are shown relative to the predicted value for each star, i.e.,    
relative to $\dot{M}_{Vink}$. For \civ, the mass-loss rates are the     
ones in our final models. For \ciii, \siiv, \niv, \pv, and \nv, {\it we remind that the   
points represent only upper limits, i.e., the mass-loss rates   
must be lower than what is displayed}.   
    
Strikingly, we see from Figure \ref{upperplot} that     
in all cases the points, i.e., upper limits     
on the mass-loss rate, fall below the theoretical line.    
The only exception is \niv\, in the case of $\zeta$ Oph,     
for which we have assigned a conservative upper limit value.    
If we consider only the \civ\, and \nv\, lines, we see that    
most results are below about -1.0 dex,   
i.e., the $\dot{M}$'s must be {\it less} than about ten times    
the values predicted by theory ! If we analyze all other    
transitions together, we see that most points fall at approximately    
$(-0.5 \pm 0.2)$ dex. This means that the mass-loss rates are    
still very low: they must be {\it less} than about two to five times    
$\dot{M}_{Vink}$! Taken together, these results bring for the  
first time additional support to the reality of weak winds.  
 
If we focus only on one object, it is also interesting  
to note in Figure \ref{upperplot} that the   
points below the highest point (i.e. closest to Vink's line)   
have the respective theoretical lines already in disagreement   
with the observed ones. In this sense, whenever we skip to models/points    
higher than others, Figure \ref{upperplot} can be interpreted    
as errors being accumulated in the model fits.     
   
We remind that all the models presented in this paper     
do not include clumping. Therefore, some of the upper limits presented     
above could be in fact much lower if the wind of    
the objects studied here are not homogeneous. In this case, we    
could then have that:     
    
\[ log \frac{\dot{M}_{clumped}}{\dot{M}_{Vink}} = log    
\frac{\dot{M}}{\dot{M}_{Vink}} + log \sqrt{f},  \]    
    
where the first term in the right-hand side is in the vertical     
axis shown in Figure \ref{upperplot}. With a typical filling factor     
$f=0.1$, some points (depending on the wind clumping sensitivity    
of the respective lines) could go down, i.e. away from the Vink's    
prediction, by 0.5 dex. We think however, that it is more    
reasonable to compare the results of homogeneous atmosphere models    
with radiative wind models that are also homogeneous, as it is    
the case in Vink et al. (2000; 2001).   
   
\begin{figure}        
\center   
\includegraphics[width=9cm,height=9cm]{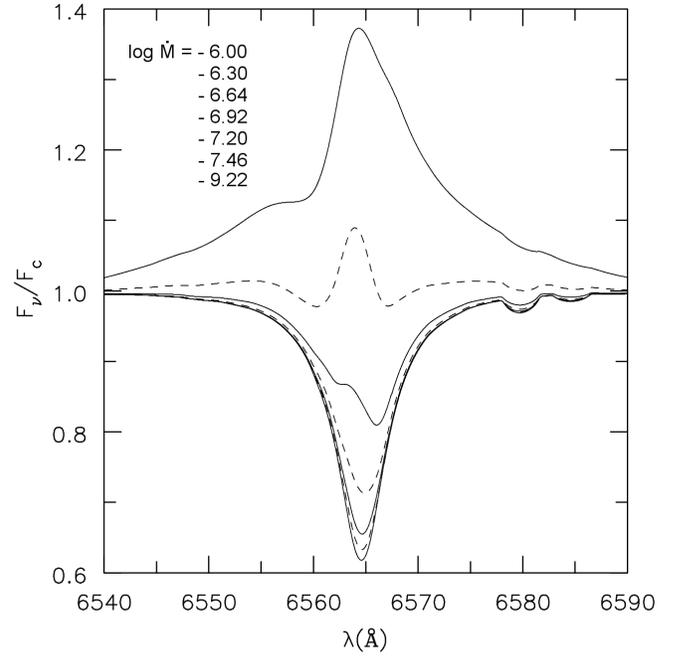}        
\caption{Sensitivity of H$\alpha$ to different values of the mass-loss   
rate. The model with the lowest mass-loss has log $\dot{M} = -9.22$ and  
it corresponds to the most intense absorption line.  The model with  
the highest mass-loss rate has log $\dot{M} = -6.0$ and it corresponds   
to the most intense line in emission. The models in between are  
shown by alternating dashed and full lines for clarity.}     
\label{halphaevol}        
\end{figure}

\subsection{H$\alpha$ behavior:}       
\label{halphasec}      
  
Although we have determined mass-loss rates from the far-UV and UV     
spectral regions, we have also analyzed the behavior of H$\alpha$     
to changes in $\dot{M}$. This is an important point because as we   
highlighted in Section \ref{summary}, the use of H$\alpha$ by   
Mokiem et al. (2005) seems to indicate for a few stars that there   
are no weak winds, i.e., the $\dot{M}$ values obtained are compatible   
with $\dot{M}_{Vink}$ (see also the discussion in Mokiem et al. 2007).  
  
Previously, the work of Martins et al. (2005) had already demonstrated 
two things regarding this line: (i) in a low   
density wind ($\dot{M} \sim 10^{-9}$ M$_{\odot}$ yr$^{-1}$) the CMFGEN   
and FASTWIND predictions agree; (ii) for $\dot{M}$ changes from   
$\sim 10^{-10}$ to $10^{-9}$ M$_{\odot}$ yr$^{-1}$ the resulting CMFGEN   
H$\alpha$ profiles are virtually identical. Here, we extend   
their analysis by presenting the changes in the H$\alpha$ profile for   
a large range in mass-loss: from $\sim 10^{-10}$ to $10^{-6}$ M$_{\odot}$  
yr$^{-1}$.  
   
\begin{figure*}        
\center   
\includegraphics[width=15cm,height=11cm]{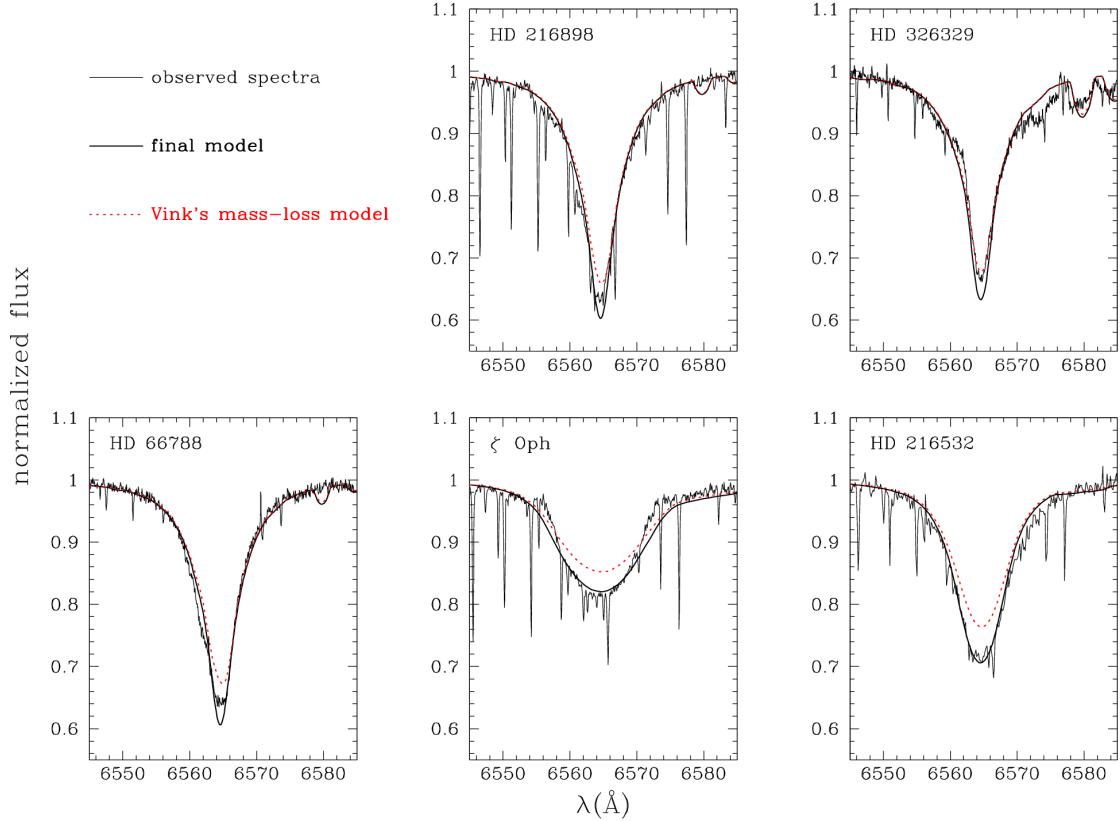}        
\caption{Fits to the observed H$\alpha$ lines in our sample using the    
mass-loss rates of our final models and the mass-loss rates predicted    
by theory ($\dot{M}_{Vink}$).}        
\label{mdot_halpha}        
\end{figure*}        
  
In Figure \ref{halphaevol} we plot models with the photospheric   
parameters fixed\footnote{In this example $vsini = 80$ km s$^{-1}$ and   
the others parameters are fixed at the values derived for one of our   
stars, namely, HD 216532.} and different values for the mass-loss rate. 
First, we note that only with a log $\dot{M} = -6.30$ the H$\alpha$ line turned to emission.  
For lower rates, the predictions are in absorption and therefore the  
measurement of $\dot{M}$ in an observed spectrum must be done based  
in the analysis of the wind filling of the photospheric profile.  
In the example shown, models with a log $\dot{M} = -9.22$, log $\dot{M} =
-7.46$, log $\dot{M} = -7.20$ do not differ much.  
Despite a factor of one hundred in $\dot{M}$, the changes in the profiles  
are too small to allow a secure discrimination of a best fit model to an observed spectrum   
(see below). Overall, the models suggest that our confidence in mass-loss rates   
derived from H$\alpha$ profiles is limited to values $\gtrsim$ 10$^{-7}$ M$_{\odot}$ yr$^{-1}$.   
As it can be seen in Figure \ref{halphaevol}, only beyond this threshold   
noticeable changes start to be found.   
   
In Figure \ref{mdot_halpha} we turn our attention to the stars of   
our sample. The fits achieved to H$\alpha$ using our final models   
(i.e. with $\dot{M}$ obtained from \civ) and models with $\dot{M}_{Vink}$   
are presented (see Table \ref{results} for the corresponding values).   
First, if we focus on HD 216898 and HD 66788, neither   
our final models nor the ones using $\dot{M}_{Vink}$ match the   
observed H$\alpha$. The models using $\dot{M}_{Vink}$ are weaker   
while our final models are stronger than the observed line center.  
However, the differences compared to the observations are very   
small: $\sim$0.02 of the continuum intensity. Given the uncertainties   
involved in the normalization of the spectra of our sample around   
H$\alpha$, we find very hard to choose with certainty between any of   
these models (and also the ones in between, i.e., with intermediate   
$\dot{M}$'s). As in Martins et al. (2005), we estimate that the position   
of the line core can have an error up to 2\%. However, if a nebular   
contamination is likely or the signal-to-noise ratio of the spectra   
is not high enough, the uncertainty increases considerably.  
Such situations illustrate that it is not straightforward to   
establish $\dot{M}$ values for O8-9V stars using H$\alpha$.   
For HD 326329 the situation is similar, but the $\dot{M}_{Vink}$   
model presents a better fit than our final model.    

Interestingly, for $\zeta$ Oph and HD 216532, our final models   
seem to be better than the ones using $\dot{M}_{Vink}$, and not only   
in the line center. To use the predicted mass-loss rates for   
these stars thus means that we can fit neither the UV   
nor the optical lines! Regarding $\zeta$ Oph,   
Repolust et al. (2004) and Mokiem et al. (2005) have presented   
$\dot{M}$ measurements from H$\alpha$ based on the FASTWIND   
code. The former authors derived an upper limit of log $\dot{M} = -6.74$
and the latter found a mass-loss rate of log $\dot{M} = -6.85$.  
In both studies, the $\beta$ parameter used for the velocity law was   
fixed at 0.8. Although the fit in Figure \ref{mdot_halpha} uses   
$\beta = 1$, we have tested a model with the same predicted   
mass-loss (log $\dot{M}_{Vink} = -6.89$) and $\beta = 0.8$.
The synthetic H$\alpha$ line gets deeper, but it is   
still not enough to fit the observations. We therefore do not confirm   
the findings of Mokiem et al. (2005). We speculate that one reason  
for this can be the different data set used and their reduction.   
Another alternative is the existence of a discrepancy between   
the predicted H$\alpha$ profile by CMFGEN and FASTWIND at   
$\sim 10^{-7}$ M$_\odot$ yr$^{-1}$. This is a possibility   
since in the comparison made by Puls et al. (2005) we   
can note that these two codes can predict slightly different   
H$\alpha$ line intensities. In the case of an O8V or a  
O10V model for example (see their Fig. 17), FASTWIND tends to present   
a stronger H$\alpha$ absorption than CMFGEN. Such differences   
can affect mass-loss rate measurements for late O dwarfs   
obtained from this line.  
 
\section{Discussion:}      
\label{discussion}   
      
It is clear from the previous section that when high   
mass-loss rates are used, i.e. above the values derived from    
\civ, a considerable discrepancy is seen between the models and   
the observations (see Figs. \ref{mdot_hd216898}-\ref{mdot_hd216532}).   
{\it Most importantly, this is seen for lines of different ions, for   
each object of our sample}. With these different diagnostics,   
the existence of weak winds gains strong additional support. 
Given the importance of this result, we now discuss the validity  
of our models. 
 
What are the consequences if we still accept the mass-loss rates 
predicted by Vink et al. (2000) for late O dwarfs as the correct ones ? 
Given our findings, we must conclude that the expanding atmosphere   
models used (from CMFGEN) lack {\it new physics}    
or have incorrect assumptions or approximations (or both). 
Even with high mass-loss rates (i.e. equal or close to $\dot{M}_{Vink}$)    
the models should provide a reasonable fit to the observed spectra,    
where only a few wind lines are seen. The current wind     
structure/ionisation then must be changed in a {\it drastic} way for a model to   
predict only a weak \civ\, wind feature (and perhaps also a weak \nv),    
despite a high $\dot{M}$ (about one hundred times higher than the ones in   
this paper or even more!). One analogy that can be done is that the  
uppermost synthetic spectrum shown in Figure \ref{sensitive_lines}  
would have to turn to the one essentially photospheric (bottom). 
Although possible, this scenario presents several difficulties. 
 
First and foremost, contrary to the recently reported case of some B supergiants, 
where a few key UV wind lines could not be reproduced by CMFGEN models (see
Searle et al. 2008 and also Crowther et al. 2006)\footnote{In
this case, a problem in the ionization structure of the models is claimed.
In fact, their wind structure seems to be distinct from the ones found in O stars (Prinja et
al. 2005). We note however, that X-rays were not used in their analysis and might 
resolve some of the discrepancies.}, we do find  
agreement with the observations (see Figs. \ref{hd216a}-\ref{hd216532b}).
Although in most spectra only one conspicuous wind signature is observed,  
the lack of wind lines such as \pv, \ciii,  
\siiv, and \niv, is also a constraint attained by our models.   
In fact, these are transitions from the photospheric region (which  
is smoothly connected to the wind) that can be well reproduced.   
 
In addition, the reasonable fits that we have achieved  
with very low mass-loss rates would have to be merely an unfortunate  
coincidence. This would be also true regarding the stars analyzed by  
Martins et al. (2005). Another issue is that depending on the physical  
ingredient(s) neglected or assumptions to be revised, the theoretical hydrodynamical   
predictions will probably need to be re-examined. In this case, it would  
now mean that we should trust neither in atmosphere nor in the current  
radiative wind models when dealing with late O dwarf stars. The weak wind problem    
would then be ill-defined. Further, any future, more sophisticated models   
would have to achieve fits to the observed spectra similar to ours,   
with perhaps more free input parameters or assumptions (e.g.   
as in non-spherical winds).  
  
There is no clear answer as for the missing ingredient(s) in the current atmosphere models.
For massive stars in general, it is true that  
although CMFGEN and other codes such as FASTWIND are considered  
{\it state-of-art}, different physical phenomena are not taken into  
account properly or are not taken into account at all due to several  
technical challenges and lack of observational constraints 
(e.g. non-stationarity; wind rotation; non-sphericity;  
realistic description of clumping; see Hillier 2008 and references therein).  
It is possible that advances in some of these topics can cast a new light  
in the analysis of massive stars. However, the consequences regarding the  
weak wind problem are difficult to foresee. 
 
{\it In short, we believe that the inclusion of new physics or relaxation  
of standard assumptions certainly deserves to be addressed in future studies  
(with the appropriate observational constraints), but our atmosphere models  
for the O8-9V stars are the best working hypothesis currently available.} 
 
Below we speculate about some possibilities worth to be investigated, and  
thereafter present the hypothesis of highly (X-rays) ionized winds. 
 
\subsection{Magnetic fields and multi-component winds} 
\label{john} 
 
From the best fit models, we can examine how closely the momentum absorbed by  
the wind matches that required to satisfy the momentum equation. In general,  
for the O8-9V stars, we find that the wind absorbs too much momentum. This can be  
illustrated by noting that the mass-loss rate from a single saturated line  
is approximately $L/c^2$ (Lucy \& Solomon 1970). In all cases,  
this value exceeds our best fit mass-loss rates (see Table \ref {results}).  
While none of the wind lines in the models are saturated, there are  
sufficient weak lines to yield a force similar, or larger, than  
the single line limit. 
 
It is unclear how this discrepancy can be removed. Possibilities include the  
existence of magnetic fields, and/or the presence of a significant component  
of hot gas. The latter could be generated by shocks in the stellar wind.  
As noted by Drew et al. (1994) and Martins et al. (2005), it is possible  
that at the densities encountered in these winds the shocked gas never cools (see also Krti\v{c}ka \& Kub\'at 2009).  
Such gas would not be easily detectable in the UV since it could be collisionally ionized  
to such an extent that very few C\,{IV} and N\,{V} ions would exist. Further  
evidence for a significant amount of hot gas comes from the observed X-ray  
fluxes; the required filling factors in the models (while strongly mass-loss 
rate dependent) indicate that the hot gas in the wind is not a trace component.
 
Of interest in this regard are the distinct profiles presented by \civ.  
While all show blue shifted absorptions, the depth and shapes  
differ significantly among the stars (see for instance the ones in HD 66788,  
HD 326329, and HD 216532). Also of concern is the general weakness of the  
P-Cygni emission components (with the possible exception of $\zeta$ Oph).  
Instead of very low mass-loss, such an absence might be related to  
complicated wind structures.  
 
\subsection{X-rays highly ionized winds:}   
\label{highlyionized} 
 
\begin{figure*}        
\centering        
\includegraphics[width=16cm,height=8cm]{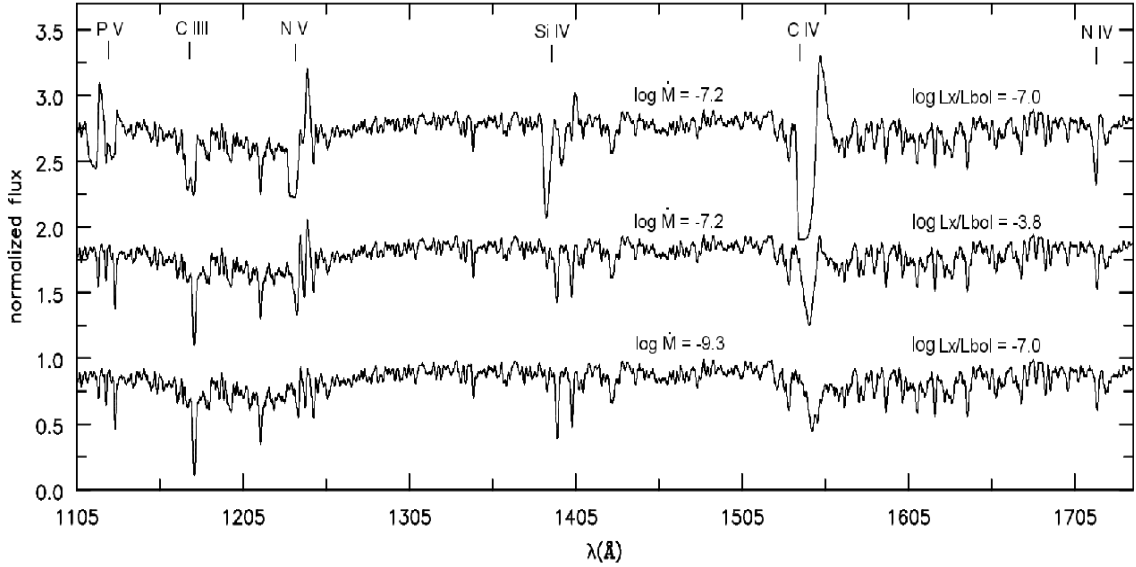}        
\caption{Effects of different log $L_X/L_{Bol}$ ratios in the ultraviolet synthetic spectrum.  
Note the similarities between the middle and bottom spectra, despite   
the very different mass-loss rates.}   
\label{xrays}        
\end{figure*}         

Although good fits could be achieved with very low $\dot{M}$'s, we have 
still studied some alternatives to have agreement with the observations 
using $\dot{M}_{Vink}$. It became soon clear that the most promising 
option was to explore high levels of X-rays emission. If most of the wind is ionized, i.e.    
is in the form of a hot plasma, we do not expect significant    
wind emission in the UV lines.   
 
In order to test the idea above, we explored models with    
high values for the $L_X/L_{Bol}$ ratio. We have found    
that indeed the wind emission is decreased in the desired    
manner. To illustrate this, we present in Figure \ref{xrays}    
two models with the same (typically Vink) mass-loss rate but with   
different values for log $L_X/L_{Bol}$: the canonical (-7.0) and a  
much higher ratio (-3.8). In addition, a third model with a  
very low mass-loss rate (log $\dot{M} = -9.3$; as in a typical final model)  
is shown with a log $L_X/L_{Bol} = -7.0$. Given the high level of X-rays emission,   
the model with a log $L_X/L_{Bol} = -3.8$ has most of its   
wind ionized and despite the high mass-loss rate, no significant line emissions    
take place, as in weak winds! Indeed, the synthetic spectrum with    
high $\dot{M}$ and log $L_X/L_{Bol}$ resembles the one with very low $\dot{M}$    
and normal log $L_X/L_{Bol}$ (see the spectra in the middle and bottom    
of Figure \ref{xrays}).  
   
In order to better quantify the result above, we have turned    
our attention to the ionisation fractions of P~{\sc v}, N~{\sc v},    
Si~{\sc iv}, and C~{\sc iv}. For the resonance lines \pv, \nv,    
\siiv, and \civ, it is well known that what is necessary to fit    
their observed profiles is an appropriate value of the product    
of the mass-loss rate and the respective ionisation fraction -    
$\dot{M} \times q_i$ - which means to find their correct/observed    
optical depths (see for instance the case of \pv\, in Fullerton et al. 2006).   
Thus, from our final models it can be said that we actually    
find\footnote{It is important to note that if the ionisation    
fractions are reliable/correct, we get $\dot{M}$ from the model    
fits. Otherwise, what is actually obtained is $\Psi _i$.}:   
   
\begin{equation}   
\dot{M} \times q_i = \Psi _i,   
\end{equation}   
   
where $\Psi _i$ is a constant that depends on the observed spectrum    
and $q_i$ is the ionisation fraction of the ion $i$. This latter is usually defined as:   
   
\begin{equation}   
q_i = \frac{\int _{0.2}^{1} n_{i}(x)dx}  { \int _{0.2}^{1} n(x)dx },   
\end{equation}   
   
where $x = v(r)/v_\infty$ and $n_{i}$ and $n$ are the ion    
and element number density, respectively. We can now compute what    
is the ionisation fraction needed to fit the observations    
by using $\dot{M}_{Vink}$, which we call $q_{i,Vink}$. Since the    
constant $\Psi _i$ gives the appropriate fit to the observations,    
we can write:   
   
\begin{equation}   
q_{i,Vink} = \Psi _i / \dot{M}_{Vink}.   
\end{equation}   
    
After we found $q_{i,Vink}$ from the equation above, for each ion,    
we have computed several models with the mass-loss rate fixed    
at $\dot{M}_{Vink}$ but with different values of log $L_X/L_{Bol}$.    
For each model, we have derived the ionisation fraction of each ion    
following Eqn. 2. We call these ionisation fractions $q_X$, i.e.,    
the ionisation fractions for a specific log $L_X/L_{Bol}$ value.    
We have then verified which one of the several $q_X$ was equal to    
$q_{i,Vink}$, in order to find for which amount of X-rays we would    
have agreement with the observations using $\dot{M}_{Vink}$.     
        
\begin{figure}        
\center   
\includegraphics[width=9cm,height=9cm]{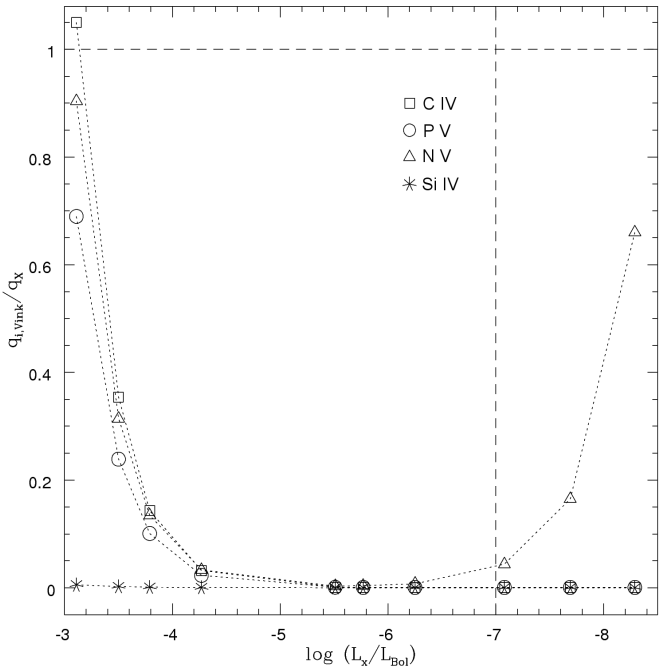}        
\caption{Ionisation fractions as a function of log $L_X/L_{Bol}$.  
The values for $q_X$ were computed using Eqn. 2 and $\dot{M}_{Vink}$.    
The $q_{i,Vink}$ values were derived from Eqn. 3 (see text for more details).}        
\label{xrays2}        
\end{figure}

In Figure \ref{xrays2} we plot the ratio $q_{i,Vink}/q_X$ for each ion    
versus several log $L_X/L_{Bol}$ values. For simplicity, we    
illustrate the case of only one object of our sample, HD 216898.   
The first thing to note is that the ions do not behave exactly    
in the same manner. This is not so surprising, since they    
have different electronic structures and ionization potentials,    
and thus react differently to X-rays. If we focus on  C~{\sc iv}    
and P~{\sc v}, we can see that only when very high log $L_X/L_{Bol}$ ratios    
are considered (around -3.0), $q_X$ gets close to $q_{i,Vink}$.    
Otherwise, their ratio is much lower than unity.    
In the case of Si~{\sc iv}, $q_{i,Vink}/q_X$ remains very    
small regardless the value of log $L_X/L_{Bol}$. However, an   
increase is observed when we use a log $L_X/L_{Bol}$ near -3.0.    
For a log $L_X/L_{Bol} \sim -7$ for example, we have    
$q_{i,Vink}/q_X \sim 5 \times 10^{-5}$. For log $L_X/L_{Bol}    
\sim -3$, this ratio is much higher, $q_{i,Vink}/q_X \sim 0.005$.    
For this ion therefore, the conclusion is that even more X-rays    
seems to be required to $q_X$ reach $q_{i,Vink}$.    
The situation for N~{\sc v} is quite    
interesting: for both low and high values of log $L_X/L_{Bol}$,    
the $q_X$ approaches $q_{i,Vink}$. This is however, not hard    
to explain.  For high values of log $L_X/L_{Bol}$, we observed that    
practically all nitrogen is concentrated in N~{\sc vi}.    
Thus, the ionisation fraction $q_X$ obtained for  N~{\sc v}    
is low as the ionisation fraction of Vink, $q_{Vink}$,    
computed from Eqn. 3. On the other hand, when very low    
log $L_X/L_{Bol}$ ratios are considered, we verified that    
most of the nitrogen is in N~{\sc iii-iv}. Thus, again,    
a very low  $q_X$ is obtained for N~{\sc v}. In    
practice, this means that the observed, weak \nv\, line (or its absence!), can    
be reproduced by models using $\dot{M}_{Vink}$ with    
either very low or large amounts of X-rays.    
   
Our conclusion from both Figures \ref{xrays} and \ref{xrays2} is    
that the only way to have high mass-loss rates ($\dot{M}_{Vink}$)    
and still agree with the observed spectra is by using a log $L_X/L_{Bol} \gtrsim -3.5$.   
Unfortunately, this is not supported by X-rays observations, where    
the log $L_X/L_{Bol}$ values measured are normally    
about -7.0 (see e.g. Sana et al. 2006) with a scatter of    
about $\pm 1.0$. It is encouraging however, to explore    
in future studies different scenarios of X-rays emission in    
stellar winds.    
 
\section{Conclusions:}     
\label{conclusions}
      
We have analyzed a sample of five late-type Galactic O dwarfs by  
using atmosphere models from the CMFGEN and TLUSTY codes   
(HD 216898, HD 326329, HD 66788, $\zeta$ Oph,   
and HD 216532). Model fits for the far-UV, UV and optical   
observed spectra were presented, from which stellar and wind   
parameters were obtained. The mass-loss rates were obtained   
first using the \civ\, feature and then we have explored   
new diagnostic lines. Our main motivation to study O8-9V   
stars was to address the so-called weak wind problem,   
recently introduced in the literature. The main findings   
of our study are summarized below:   
   
\begin{itemize}   
   
\item[$\triangleright$] The stellar parameters obtained for our sample   
are quite homogeneous (see Table \ref{results}). Surface gravities of about 3.8   
(after correcting for $vsini$) to 4.0 and $T_{eff}$'s of   
30 to 34$kK$ were obtained. These values show a good agreement   
with the latest calibrations of Galactic O star parameters,   
regarding the spectral types of the programme stars.\\   
   
\item[$\triangleright$] By using the \civ\, line we have derived mass-loss   
rates considerably lower than theoretical predictions   
(Vink et al. 2000). A discrepancy of roughly two orders of   
magnitude is observed (see Figure \ref{results_plot}). We thus confirm the results of   
the study of Martins et al. (2005) of weak winds   
among Galactic O8-9 dwarf stars. We also confirm a breakdown   
or a steepening of the modified wind momentum luminosity   
relation for low luminosity objects (log $L_\star/L_\odot \lesssim 5.2$).\\   
   
\item[$\triangleright$] We have investigated the carbon abundance    
based on a set of UV (\ciii) and optical photospheric lines.  
We estimated that the following    
range is reasonable for our programme stars: $0.5 \lesssim \epsilon _C / \epsilon   
_{C_{\odot}} \lesssim 2$ (in mass fractions). Although the uncertainty    
is large, it is not large enough to be the reason of the weak wind problem.\\   
   
\item[$\triangleright$] We have explored different ways (besides using   
\civ) to determine mass-loss rates for late O dwarfs.    
We have found that \pv, \ciii, \nv, \siiv, and \niv\,   
lines are good diagnostic tools. We have used each   
of them to derive independent mass-loss rate limits,   
for each star of our sample. We have found that together   
with \civ, the use of the \nv\, line implies in the lowest    
$\dot{M}$ upper limit rates. The results obtained show that the   
$\dot{M}$ values must be {\it less} than about -1.0 dex   
compared to $\dot{M}_{Vink}$. By considering the   
other lines, we still find very low mass-loss rates.   
The results obtained show that $\dot{M}$ must be   
{\it less} than about $(-0.5 \pm 0.2)$ dex compared to    
$\dot{M}_{Vink}$. They bring additional support to the    
reality of weak winds.\\   
  
\item[$\triangleright$] Upper mass-loss rate limits derived from \niv\, 
and \ciii\, can be a factor of three, or more, lower than the 
theoretical mass-loss rates of Vink. This is crucial in confirming a mass-loss  
discrepancy as \niv\, and to a lesser extent \ciii\,   
are formed close to the photosphere, where the uncertainties in  
the ionization structure should be much less than in the (outer) wind.\\ 
  
\item[$\triangleright$] We have analyzed the H$\alpha$ line and    
we observed that its profile is insensitive to $\dot{M}$    
changes when we are in the $< 10^{-8}$M$_\odot$ yr$^{-1}$ regime.    
For some objects of our sample it is uncertain to choose between  
the H$\alpha$ fit presented by our final models and the ones  
using $\dot{M}_{Vink}$. The interpretation of the results  
is hindered by uncertainties in the continuum normalization   
of echelle spectra. Such situation gets even more complicated   
if a contamination by nebular emission is likely. For   
the stars HD 216532 and $\zeta$ Oph, the fits to the observed    
H$\alpha$ lines using $\dot{M}_{Vink}$ were not satisfactory.    
This result shows that even when H$\alpha$ is used, lower     
than predicted mass-loss rates can be preferred.\\   
  
\item[$\triangleright$] We have investigated ways to have    
agreement with the observed spectra of the O8-9V stars    
using the mass-loss rates predicted by theory ($\dot{M}_{Vink}$).    
The only mechanism plausible that we found is X-rays.    
By using high values for the log $L_X/L_{Bol}$ ratio, we could observe that    
very few wind emissions takes place, as in models with very low    
mass-loss rates ($\sim 10^{-10} - 10^{-9}$ M$_\odot$ yr$^{-1}$).    
However, the values needed to be used (log $L_X/L_{Bol} \gtrsim -3.5$)    
are not supported by the observations, which usually measure    
log $L_X/L_{Bol}$ values near -7.0.   
   
\end{itemize}   
 
Although our analysis was performed with {\it state-of-art}    
atmosphere models, there are a couple of issues which still    
need to be addressed in future studies. For example,    
some stars of our sample (HD 326329 and HD 216532) present a    
deep \civ\, feature that could not be well reproduced. The origin of 
this extra absorption remains to be investigated (see Section \ref{john}). 
Studies concerning non-spherical winds, specially for stars that are fast  
rotators, are also of great interest. First synthetic spectra computed by our group have  
shown interesting line profile changes. A deeper analysis    
is needed to see if some of the discrepancies found in the    
case of $\zeta$ Oph can be solved. Furthermore, a more  
realistic treatment or even alternative descriptions of the X-rays  
emission in the atmosphere models (e.g. in conformity with  
the magnetic confinement scenario; Ud-Doula \& Owocki 2002) 
could bring valuable informations. First steps in this direction  
have been taken by Zsarg\'o et al. (2009). As we have shown in Section   
\ref{highlyionized}, X-rays are an efficient mechanism to   
change the ionisation structure of the stellar winds.   
   
It would be also very useful to analyze a sample of O stars    
having log $L_\star/L_\odot$ near $\sim 5.2$. The disagreement    
between atmosphere models and theoretical predictions for Galactic    
stars seems to start around this value. The candidate spectral types       
would be O6.5V, O7V, and O7.5V. Also, this same question   
needs to be better studied in metal-poor environments, i.e.   
in the LMC and SMC. We intend to investigate these issues    
in a future paper.   
  
From the point of view of the hydrodynamics, there are also important   
questions that could be addressed. In the work of Vink et al.   
(1999; 2000; 2001), a predicted mass-loss rate is computed   
based on a set of ISA-WIND models (de Koter et al. 1997) plus  
the use of a Monte-Carlo technique to compute the radiative acceleration   
($g_L$). Although the results obtained by their procedure are self-consistent,   
it should be kept in mind that the ISA-WIND atmosphere models do not include  
the effect of X-rays. As the wind ionization can be changed significantly,  
new radiative accelerations can be found and perhaps new predicted mass-loss  
rates might be derived. Within this picture, the agreement with the Vink et al. predictions   
for early O dwarfs may be understood, since their winds are not seriously  
affected by X-rays. 
  
Our results bring new constraints to the weak wind problem.  
Galactic O8-9V stars seem to present very low mass-loss rates,  
as indicated by the \civ\, feature, and also additional  
spectral diagnostics. The existence of weak winds poses a  
challenge to the current radiative wind models, and possibly  
has important consequences to the massive stellar evolution theory.   
  
        
\begin{acknowledgements}        
        
W. M. acknowledges the travel grant provided by IAU (Exchange of Astronomers    
Programme) and CNES for the postdoctoral fellowship. J-CB acknowledges   
financial support from the French National Research Agency (ANR) through   
program number ANR-06-BLAN-0105. TL and DJH were supported by the NASA   
Astrophysics Data Program (grant NNG04GC81G). We wish to thank Ya\"el Naz\'e for
helping with the normalization of the spectrum of HD 216898. We also wish to thank 
an anonymous referee for useful comments which helped to improve the paper. 
This research has made use of the SIMBAD database, operated at CDS, Strasbourg, France.
        
\end{acknowledgements}        
        
{}


\Online

\begin{appendix}        
\section{$\dot{M}$ upper limits:}      
In this Section, we present the model fits used to establish upper limits on the
mass-loss rate from different UV lines.

\begin{landscape}
\begin{figure}[!h]
\center    
\includegraphics[width=25cm,height=17cm]{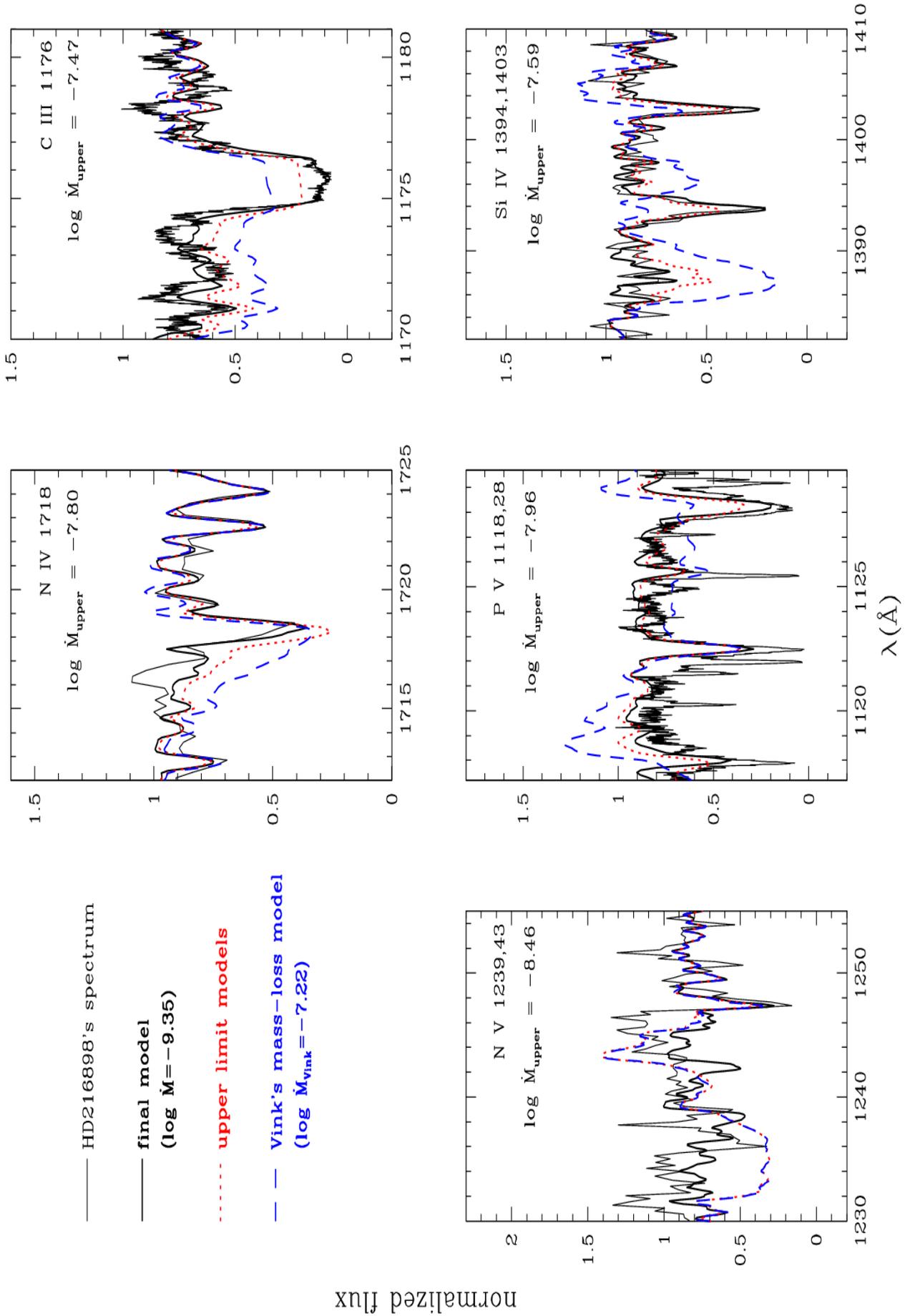}    
\caption{Upper limits for the mass-loss rate in HD 216898.}
\label{mdot_hd216898}        
\end{figure}
\end{landscape}
     
\begin{landscape}
\begin{figure}[!h]     
\center   
\includegraphics[width=25cm,height=17cm]{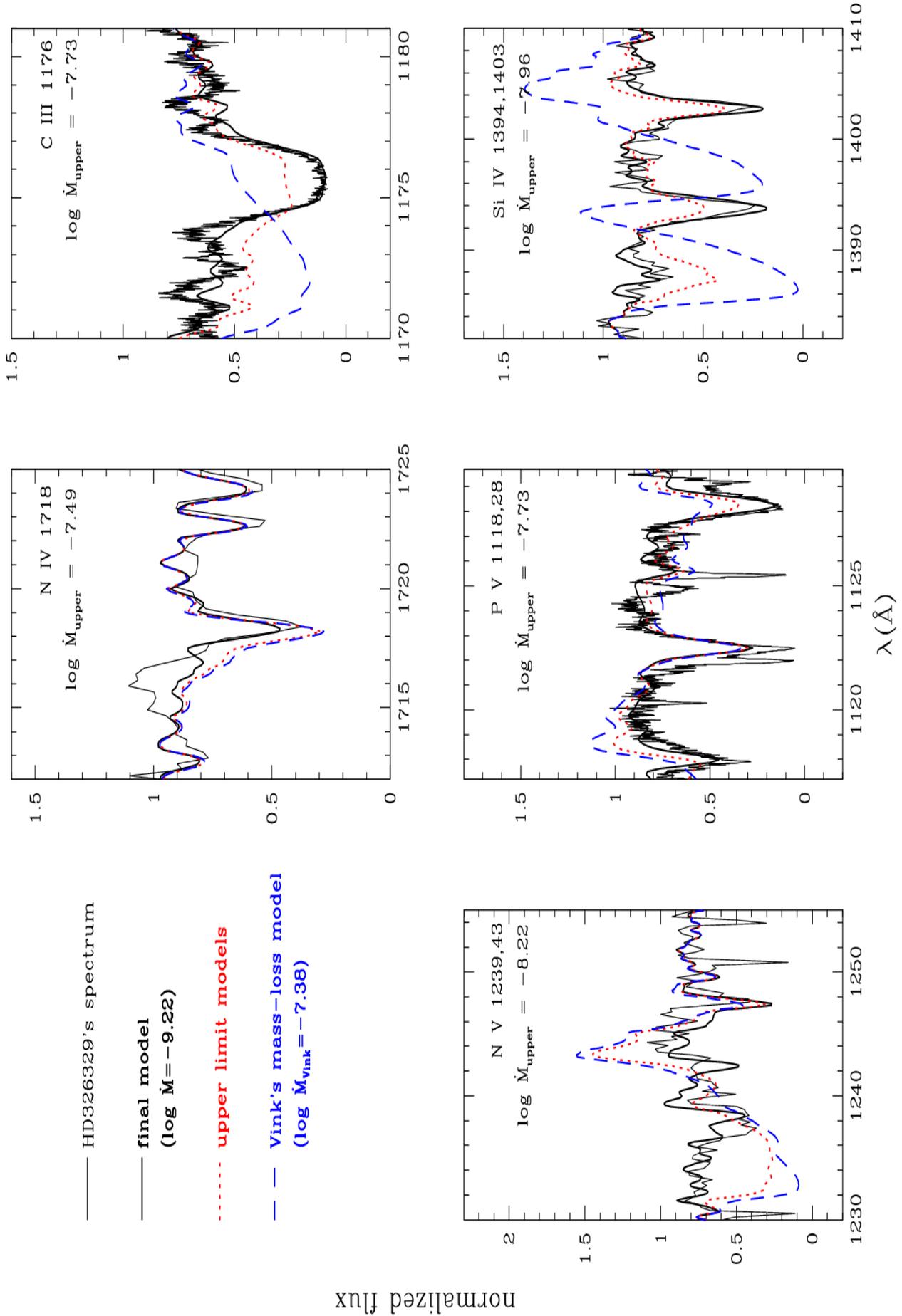}   
\caption{Upper limits for the mass-loss rate in  HD326329.}        
\label{mdot_hd326329}        
\end{figure}
\end{landscape}

\begin{landscape}
\begin{figure}[!h]     
\center   
\includegraphics[width=25cm,height=17cm]{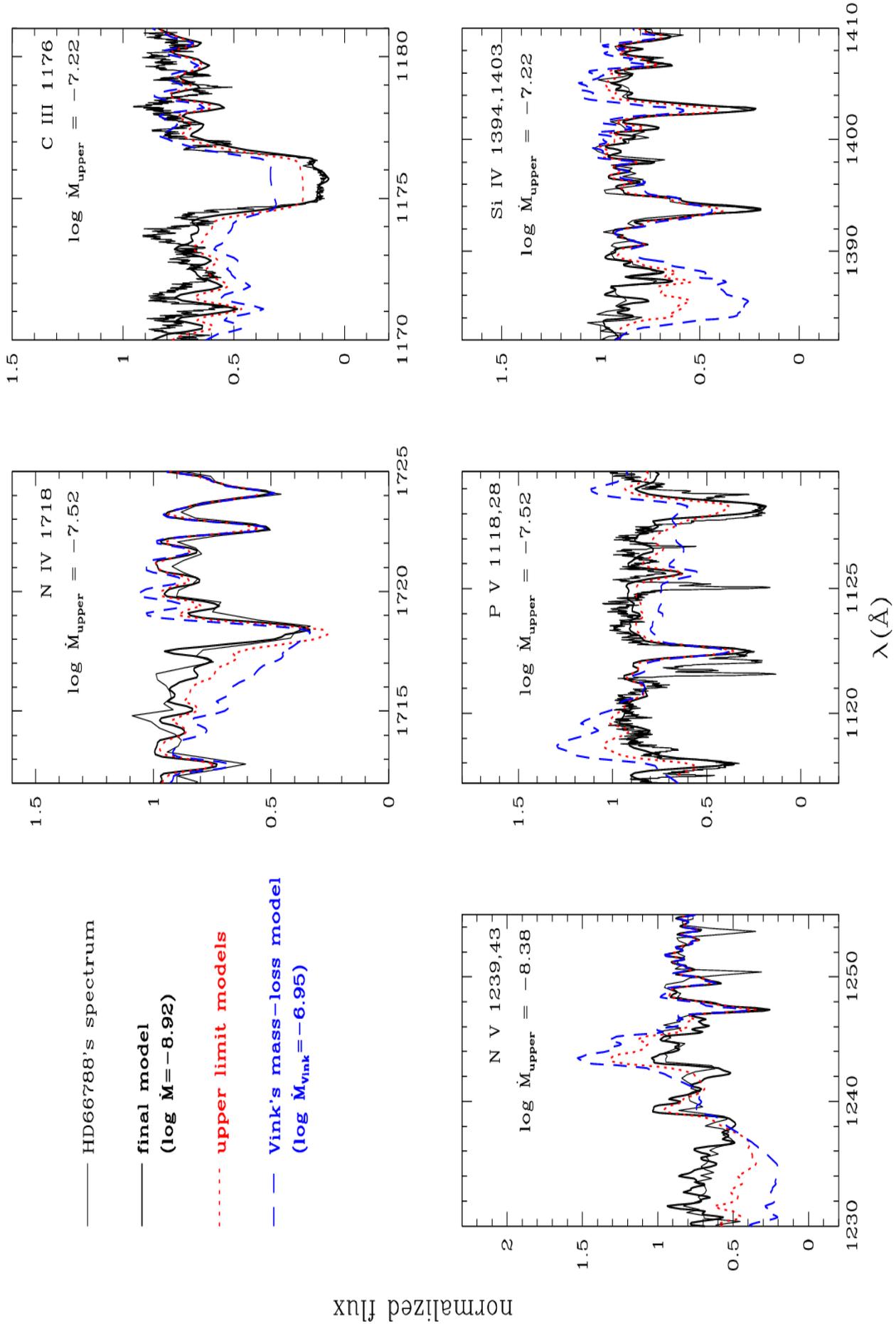}   
\caption{Upper limits for the mass-loss rate in HD66788.}        
\label{mdot_hd66788}        
\end{figure}
\end{landscape}

\begin{landscape}          
\begin{figure}[!h]        
\center   
\includegraphics[width=25cm,height=17cm]{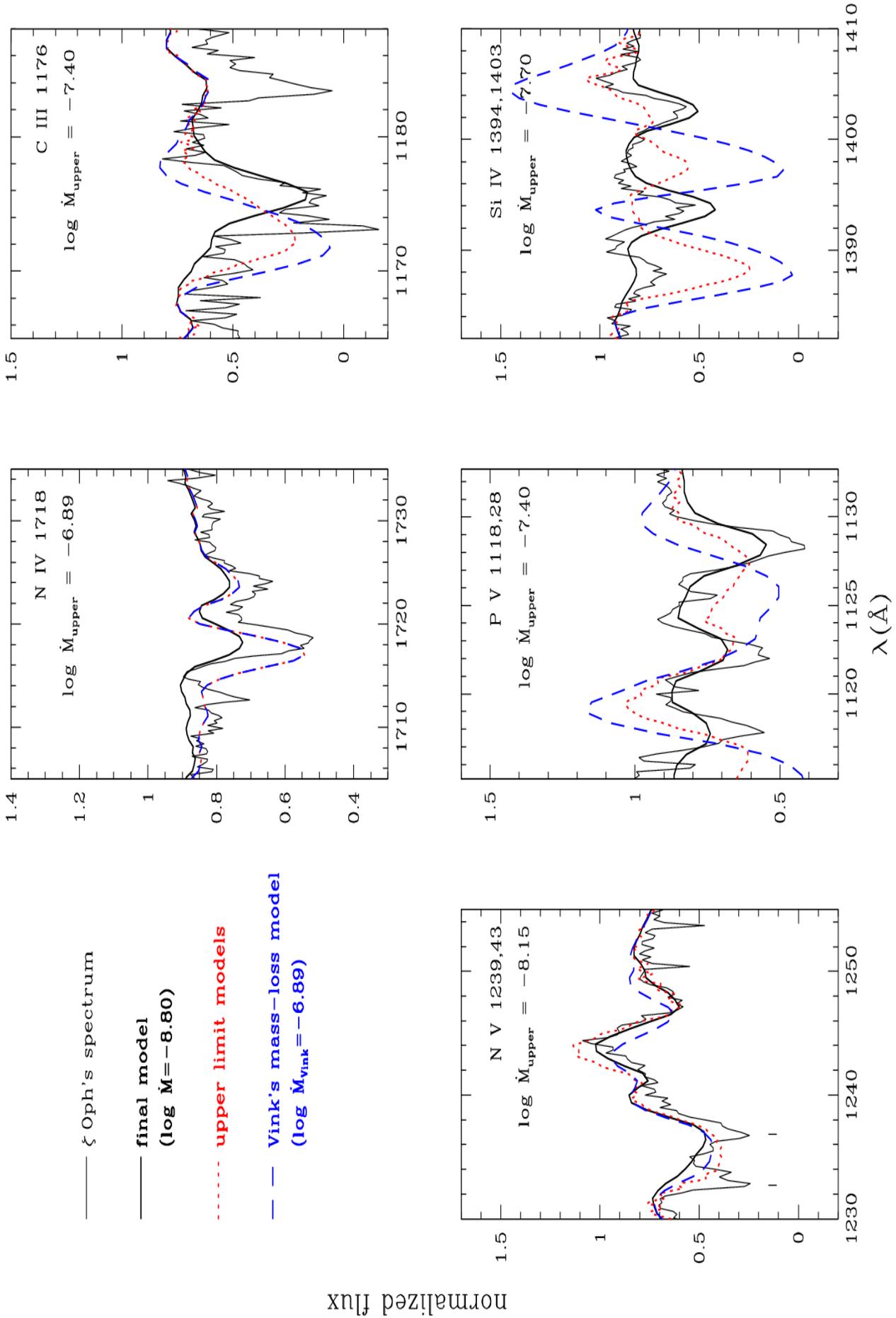}   
\caption{Upper limits for the mass-loss rate in $\zeta$ Oph.}        
\label{mdot_zeta}        
\end{figure}    
\end{landscape}

\begin{landscape}  
\begin{figure}[!h]        
\center   
\includegraphics[width=25cm,height=17cm]{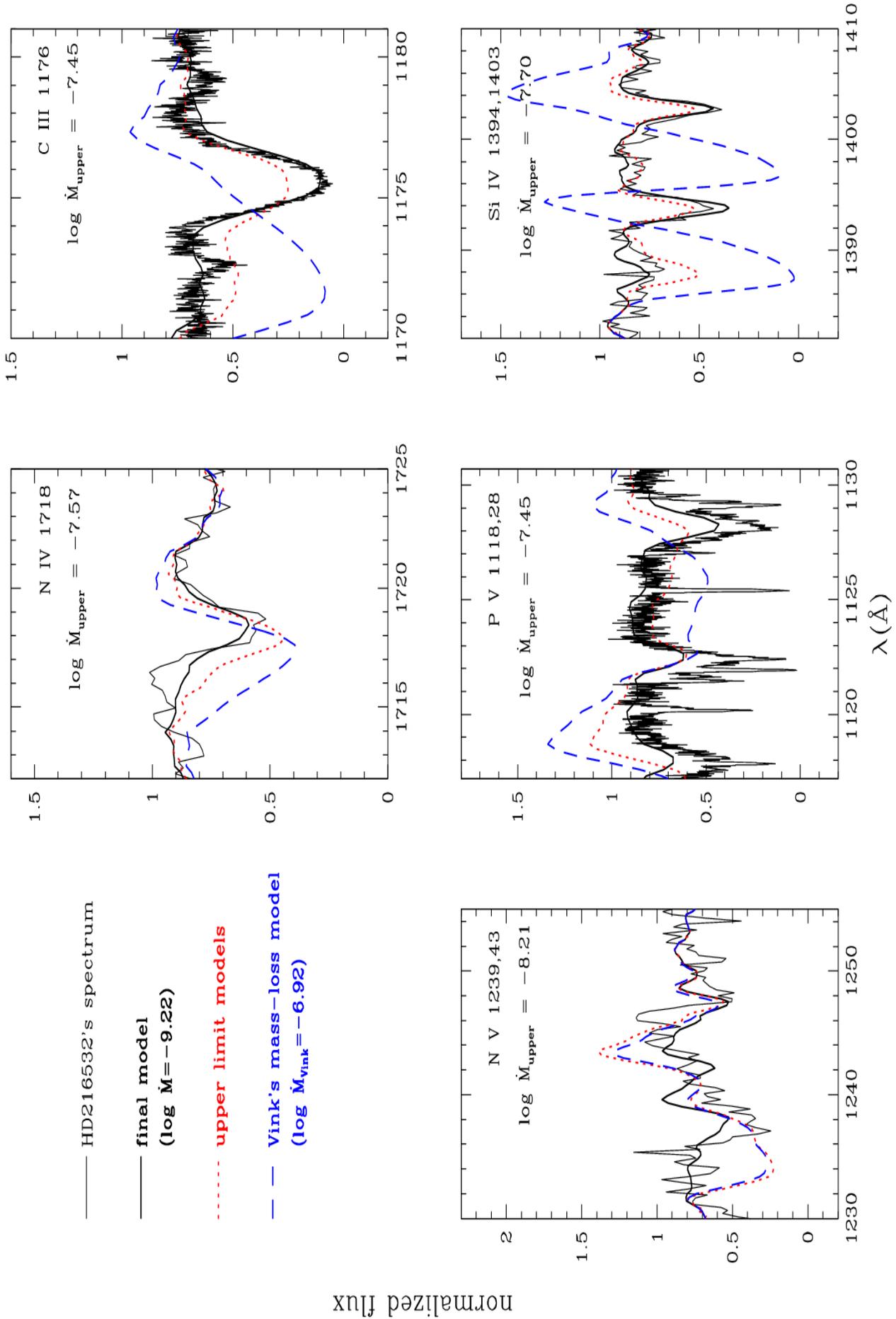}     
\caption{Upper limits for the mass-loss rate in HD 216532.}        
\label{mdot_hd216532}        
\end{figure}
\end{landscape}
 
\end{appendix}

 \end{document}